\documentclass[12pt, draftclsnofoot, onecolumn]{IEEEtran}

\usepackage{amsmath}
\usepackage{amssymb}
\usepackage{amsfonts}
\usepackage{graphicx}
\usepackage{epsfig}
\usepackage{subfigure}
\usepackage{psfrag}
\usepackage{cite}
\usepackage{latexsym}
\usepackage{url}
\usepackage{color}
\usepackage{multirow}

\PassOptionsToPackage{bookmarks={false}}{hyperref}

\begin{document}
\title{UAV-Enabled Wireless Power Transfer: Trajectory Design and Energy Optimization
\author{Jie Xu,~\IEEEmembership{Member,~IEEE}, Yong Zeng,~\IEEEmembership{Member,~IEEE}, and Rui Zhang,~\IEEEmembership{Fellow,~IEEE}}
\thanks{Part of this paper will be presented in the IEEE Global Communications Conference (Globecom) Workshop, Singapore, December 4-8, 2017 \cite{Conference1}, and the Asia-Pacific Conference on Communications (APCC) Workshop, Perth, Australia, December 11-13, 2017 \cite{Conference2}.}
\thanks{J. Xu is with the School of Information Engineering, Guangdong University of Technology (e-mail: jiexu@gdut.edu.cn).}
\thanks{Y. Zeng and R. Zhang are with the Department of Electrical and Computer Engineering, National University of Singapore (e-mail: \{elezeng, elezhang\}@nus.edu.sg).}
}



\maketitle
\vspace{-2em}
\begin{abstract}
This paper studies a new unmanned aerial vehicle (UAV)-enabled wireless power transfer (WPT) system, where a UAV-mounted mobile energy transmitter (ET) is dispatched to deliver wireless energy to a set of energy receivers (ERs) at known locations on the ground. We investigate how the UAV should optimally exploit its mobility via trajectory design to maximize the amount of energy transferred to all ERs during a finite charging period. First, we consider the maximization of the sum energy received by all ERs by optimizing the UAV's trajectory subject to its maximum speed constraint. Although this problem is non-convex, we obtain its optimal solution, which shows that the UAV should hover at one single fixed location during the whole charging period. However, the sum-energy maximization incurs a ``near-far'' fairness issue, where the received energy by the ERs varies significantly with their distances to the UAV's optimal hovering location. To overcome this issue, we consider a different problem to maximize the minimum received energy among all ERs, which, however, is more challenging to solve than the sum-energy maximization. To tackle this problem, we first consider an ideal case by ignoring the UAV's maximum speed constraint, and show that the relaxed problem can be optimally solved via the Lagrange dual method. The obtained trajectory solution implies that the UAV should hover over a set of fixed locations with optimal hovering time allocations among them. Then, for the general case with the UAV's maximum speed constraint considered, we propose a new {\emph{successive hover-and-fly}} trajectory motivated by the optimal trajectory in the ideal case, and obtain efficient trajectory designs by applying the successive convex programing (SCP) optimization technique. Finally, numerical results are provided to evaluate the performance of the proposed designs under different setups, as compared to other benchmark schemes.
\end{abstract}
\vspace{0em}

\newtheorem{definition}{\underline{Definition}}[section]
\newtheorem{fact}{Fact}
\newtheorem{assumption}{Assumption}
\newtheorem{theorem}{\underline{Theorem}}[section]
\newtheorem{lemma}{\underline{Lemma}}[section]
\newtheorem{corollary}{\underline{Corollary}}[section]
\newtheorem{proposition}{\underline{Proposition}}[section]
\newtheorem{example}{\underline{Example}}[section]
\newtheorem{remark}{\underline{Remark}}[section]
\newtheorem{algorithm}{\underline{Algorithm}}[section]
\newcommand{\mv}[1]{\mbox{\boldmath{$ #1 $}}}
\setlength\abovedisplayskip{4pt}
\setlength\belowdisplayskip{4pt}

\vspace{-1em}

\section{Introduction}

Radio frequency (RF) transmission enabled wireless power transfer (WPT) is a promising solution to provide perpetual and cost-effective energy supplies to low-power electronic devices, and it is anticipated to have abundant applications in future Internet-of-things (IoT) wireless networks (see, e.g., \cite{BiHoZhang2015,ZengZhang2017} and the references therein). In conventional WPT systems, dedicated energy transmitters (ETs) are usually deployed at fixed locations to send RF signals to charge distributed energy receivers (ERs) \cite{BiZhang2016} such as low-power sensors or IoT devices. However, due to the severe propagation loss of RF signals over long distance, the performance of practical WPT systems for wide coverage range is fundamentally constrained by the low end-to-end power transmission efficiency. As a consequence, in order to provide ubiquitous wireless energy accessibility for massive low-power ERs distributed in a large area, fixed-location ETs need to be deployed in an ultra-dense manner. This, however, would tremendously increase the cost, and thus hinder the large-scale implementation of future WPT systems. In the literature, various approaches have been proposed aiming to alleviate this issue by enhancing the WPT efficiency at the {\it link level}, including multi-antenna energy beamforming \cite{XuLiuZhang2014,XuZhang2014,XuZhang2016,ZengZhang2015,Schober2016,Yang2014,Larsson2016}, energy scheduling \cite{BiZhang2016b,Aissa2015}, and energy waveform optimization \cite{ClerckxBayguzina2016,Moghadam2017}. Different from these prior studies, in this paper we tackle this problem from a fundamentally new perspective at the {\it system level}, i.e., we propose a radically novel architecture for WPT systems by utilizing unmanned aerial vehicles (UAVs) as mobile ETs.

UAV has drawn significant research interests recently due to its wide range of applications, including surveillance and monitoring, aerial radar and camera, cargo delivery, communication platforms, etc. Particularly, mounted with miniaturized communication transceivers, low-altitude UAVs can be used as aerial mobile base stations (BSs) or relays to help enhance the performance of terrestrial wireless communication systems (see, e.g., \cite{ZengZhangLim2016} and the references therein). By optimizing the UAV's trajectory jointly with communication scheduling, the air-to-ground link distances between the UAV and its served ground users can be effectively shortened, thus significantly improving the system throughput \cite{ZengZhangLim2016b,Zeng17}.

\begin{figure}
\centering
 \epsfxsize=1\linewidth
    \includegraphics[width=9cm]{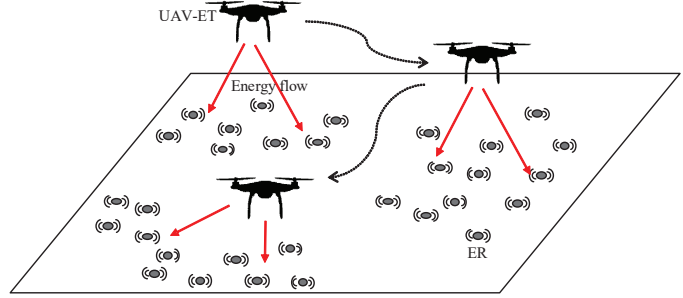}
\caption{Illustration of a UAV-enabled WPT system.} \label{fig:UAV:WPT}\vspace{-2em}
\end{figure}

Motivated by UAV-assisted wireless communications, in this paper we propose a new UAV-enabled WPT architecture as illustrated in Fig.~\ref{fig:UAV:WPT}. With the proposed architecture, a group of UAVs are dispatched as mobile ETs that fly above the serving area to cooperatively charge a set of distributed ERs on the ground. We assume that these ERs have fixed locations that are {\it a priori} known for the UAV trajectory design. By exploiting the fully controllable mobility of the UAVs via trajectory design, the proposed system is expected to significantly improve the WPT performance, while reducing the number of required ETs as compared to the conventional WPT systems with ETs deployed at fixed locations. Notice that there have been some prior works (e.g., \cite{Xie2012,Shu2016}) that considered the use of moving ground vehicles as mobile charging stations to wirelessly charge sensor nodes. Different from ground vehicles that have only limited mobility in a two-dimensional (2D) area usually with a large number of ground obstacles, UAVs can be more flexibly deployed and moved in the three-dimensional (3D) free space. Furthermore, compared to terrestrial wireless channels that typically suffer from various impairments such as shadowing and fading in addition to path loss, UAVs usually possess better channels to ground ERs due to the higher chance of having moderate-distance line-of-sight (LOS) links with them.

A fundamental question to be addressed for the proposed UAV-enabled WPT systems is as follows: how to jointly optimize the trajectories of multiple UAVs so as to maximize the energy transferred to all ERs in a fair manner? This question, however, has not yet been studied in the literature to our best knowledge, and it is non-trivial to be addressed, even for the simplest scenario with one UAV and two ERs \cite{Conference1}. Notice that in this basic setup, the transferred power from the UAV to the two ERs critically depends on the UAV's trajectory or locations at different time. For example, when the UAV moves from one ER to the other, their received power will decrease and increase, respectively, thus resulting in an interesting power trade-off between them.

For the purpose of exposition, in this paper we focus on the UAV-enabled multiuser WPT system with one UAV/ET and $K > 1$ ERs, while leaving the general scenario with more than one UAVs/ETs for our future work. Under the considered setup, we aim to find the optimal UAV trajectory to maximize the amount of energy transferred to the $K$ ERs during a finite charging period, subject to the UAV's maximum speed constraint. To our best knowledge, this work is the first that explores the UAV's trajectory design for WPT performance optimization. The main results of this paper are summarized as follows.

First, we consider the UAV's trajectory optimization in the horizontal plane with a fixed altitude above the ground to maximize the sum received energy of all ERs. Despite that this problem is non-convex and involves an infinite number of variables, we derive its optimal solution, which shows that the UAV should hover at one single fixed optimal location during the whole charging period, and the optimal hovering location can be obtained via a 2D exhaustive search. In particular, we obtain the optimal hovering location in closed-form for the special case with $K = 2$ ERs. It is shown that when the distance between the two ERs is smaller than a certain threshold, the optimal hovering location is exactly above the middle point between the two ERs; whereas when their distance is larger than the threshold, the optimal hovering location is closer to one ER than the other. In general, the sum-energy maximization with the optimal fixed hovering location incurs a severe ``near-far'' fairness issue, especially for a network spanning over a large area, as the near ERs in close proximity to the UAV can receive significantly more energy than the far-away ERs.

Next, to resolve the fairness issue, we consider an alternative problem to maximize the minimum received energy among all ERs via trajectory optimization. This problem is more challenging to solve than the previous sum-energy maximization. To obtain useful insight and a performance upper bound, we first consider an ideal case by assuming that the UAV maximum speed constraint can be ignored.\footnote{Notice that this assumption holds approximately if the charging duration and/or the UAV maximum speed are sufficiently large.} In this case, the problem is shown to satisfy the so-called time-sharing condition in \cite{YuLui2006}, and thus can be optimally solved via the Lagrange dual method. The obtained optimal solution reveals that the UAV should hover over an optimal set of fixed locations, with optimal hovering time allocations among them. For the special case of $K=2$ ERs, the closed-form solution is also obtained and compared against that for sum-energy maximization.

Last, we consider the above min-energy maximization problem for the general case with the UAV's maximum speed constraint considered. Inspired by the optimal multi-location-hovering solution in the ideal case, we propose a {\it successive hover-and-fly} trajectory design, where the UAV successively hovers at a given set of hovering locations (e.g., using the optimal set of hovering locations obtained in the ideal case if the charging duration is sufficiently large) each for a certain duration, and flies with the maximum speed between these hovering locations. The total flying time is minimized by finding the path with the shortest traveling distance to visit all of these hovering locations. The proposed trajectory is proved to be optimal for the case of $K = 2$, and also asymptotically optimal for $K > 2$ if the charging duration is sufficiently large so that the total flying time becomes asymptotically negligible. Furthermore, we also propose a successive convex programming (SCP)-based algorithm to obtain a locally optimal solution to the min-energy maximization problem with $K > 2$. By employing the successive fly-and-hover trajectory as the initial point, the SCP-based algorithm iteratively refines the UAV trajectory to improve the max-min energy of all ERs until convergence.

It is worth noting that in the preliminary work \cite{Conference1}, we considered the UAV-enabled WPT system in the case with $K=2$ ERs and characterized the achievable region of the received energy by the two ERs via UAV trajectory optimization; while in \cite{Conference2}, we studied the min-energy maximization for the case with $K > 2$ ERs. Different from the above two prior studies, this paper provides a more comprehensive study on both the sum-energy and min-energy maximization problems for the general case with $K \ge 2$ ERs.

The remainder of this paper is organized as follows. Section \ref{sec:system} presents the system model of the UAV-enabled WPT system. Section \ref{sec:max:sum} presents the optimal solution to the sum-energy maximization problem. Section \ref{sec:max:min} presents the optimal solution to the min-energy maximization problem when the UAV maximum speed constraint is ignored. Section \ref{sec:max:min:2} presents the proposed solutions to the min-energy maximization problem with the maximum UAV speed constraint considered. Section \ref{sec:numerical} provides numerical results to validate the effectiveness of our proposed trajectory designs. Finally, Section \ref{sec:conclusion} concludes this paper.

\section{System Model}\label{sec:system}

We consider a UAV-enabled multiuser WPT system, where a UAV is dispatched to deliver wireless energy to $K \ge 2$ ERs located on the ground. Let $\mathcal K \triangleq \{1,\ldots,K\}$ denote the set of ERs. Each ER $k \in\mathcal{K}$ has a fixed location on the ground, denoted by $(x_k,y_k,0)$ in a 3D Euclidean coordinate, which is assumed to be known to the UAV {\it{a priori}} for its trajectory design. We consider a finite charging period with duration $T$, denoted by $\mathcal{T} \triangleq [0,T]$. At each time instant $t \in \mathcal T$, the UAV is assumed to fly at a fixed altitude $H > 0$ above the ground, whose time-varying location is denoted as $(x(t),y(t),H)$. We assume that the initial and final UAV locations at time $t=0$ and $t=T$ are not pre-determined, but can be freely optimized.\footnote{This assumption is made simplify for the purpose of exposition, while our proposed algorithms in Section \ref{sec:max:min:2} can be easily modified to incorporate the initial and/or final location constraints.} Denote by $V$ in meter/second (m/s) the maximum possible speed of the UAV. We then have the maximum speed constraint at each time instant expressed as
\begin{align}\label{eqn:UAV:speed:constraint}
\sqrt{\dot{x}^2(t) + \dot{y}^2(t)} \le V, \forall t\in\mathcal T,
\end{align}
with $\dot{x}(t)$ and $\dot{y}(t)$ denoting the time-derivatives of $x(t)$ and $y(t)$, respectively.

We assume that the wireless channel between the UAV and each ER is LOS-dominated, so that the free-space path loss model similarly as in  \cite{ZengZhangLim2016,ZengZhangLim2016b} is adopted. At time $t \in \mathcal T$, the channel power gain from the UAV to ER $k \in \mathcal{K}$ is modeled as $h_k(t) = \beta_0 d_k^{-2}(t)$, where $d_k(t) = \sqrt{(x(t) - x_k)^2 +(y(t) - y_k)^2 + H^2}$ is their distance and $\beta_0$ denotes the channel power gain at a reference distance of $d_0 = 1$ m. Assuming that the UAV has a constant transmit power $P$, the received RF power by ER $k$ at time $t$ is thus given by
\begin{align}\label{eqn:harvested:power}
Q_k(x(t),y(t)) = h_k(t)P = \frac{\beta_0P}{(x(t) - x_k)^2 + (y(t) - y_k)^2 + H^2}.
\end{align}

The total energy received by each ER $k \in \mathcal{K}$ over the whole charging period is a function of the UAV's trajectory $\{x(t),y(t)\}$, which can be written as
\begin{align}\label{eqn:harvested:energy}
E_k(\{x(t),y(t)\}) = \int_{0}^T Q_k(x(t),y(t)) \text{d}t.
\end{align}

Note that at each ER, the received RF signals are converted into direct current (DC) signals for energy harvesting via rectifiers \cite{XuZhang2014}. In practice, the RF-to-DC conversion is generally non-linear and the conversion efficiency critically depends on the received RF power and waveform at the ER (see, e.g., \cite{ZengZhang2017,ClerckxBayguzina2016,Ng2015}). To the authors' best knowledge, a generic model to accurately characterize the non-linear RF-to-DC conversion efficiency is not available yet, but in general the harvested DC power monotonically increases with the received RF power. Therefore, for simplicity, in this paper we consider the received RF power in \eqref{eqn:harvested:power} and the resultant received energy in \eqref{eqn:harvested:energy} by ERs prior to RF-to-DC conversion as the performance metrics.


%
%
%

\section{Sum-Energy Maximization}\label{sec:max:sum}

In this section, we consider the maximization of the sum received energy of all ERs over the charging period, by optimizing the UAV's trajectory $\{x(t),y(t)\}$ subject to the speed constraints in \eqref{eqn:UAV:speed:constraint}. The problem can be expressed as
\begin{align}
\text{(P1)}:\max_{\{x(t),y(t)\}}  ~& \sum_{k\in\mathcal{K}} E_k(\{x(t),y(t)\}) \nonumber\\
{\text{s.t.}}~~~& \eqref{eqn:UAV:speed:constraint}. \nonumber
\end{align}
Problem (P1) involves an infinite number of optimization variables, i.e., $x(t)$'s and $y(t)$'s over continuous time $t$. Furthermore, (P1) is a non-convex optimization problem as the objective function is a non-concave function with respect to the trajectory $\{x(t),y(t)\}$. Therefore, (P1) is generally difficult to be optimally solved. In the following, we first present the optimal solution to (P1) for the general case with arbitrary value of $K\ge 2$ in Section \ref{sec:III:A}, and then consider (P1) for the special case with $K=2$ in Section \ref{sec:sum:energy:K2} to draw more insights.

\subsection{Optimal Solution to Problem (P1)}\label{sec:III:A}
For the ease of description, given the UAV's location $x(t)$ and $y(t)$ at a given time $t$, we define the sum-power received by all the $K$ ERs as
\begin{align}\label{eqn:psi}
\psi(x(t), y(t)) \triangleq \sum_{k\in\mathcal K} Q_k(x(t),y(t)) = \sum_{k\in\mathcal K} \frac{\beta_0P}{(x(t) - x_k)^2 + (y(t) - y_k)^2 + H^2}.
\end{align}
Accordingly, the sum-energy received by the $K$ ERs over the whole charging period is
\begin{align}\label{eqn:sum:energy}
\sum_{k\in\mathcal{K}} E_k(\{x(t),y(t)\}) = \int_{0}^T \psi(x(t), y(t)) \text{d}t.
\end{align}

Let $x^\star$ and $y^\star$ denote an optimal UAV location that maximizes the function $\psi(x, y)$, i.e.,
\begin{align}\label{eqn:x_star:y_star}
(x^\star,y^\star) = \arg \max_{x,y}~\psi(x, y).
\end{align}
As the function $\psi(x, y)$ is non-concave with respect to $x$ and $y$, it is generally difficult to find the closed-form expression of $x^\star$ and $y^\star$. Fortunately, $\psi(x, y)$ in problem \eqref{eqn:x_star:y_star} only has two variables $x$ and $y$. Besides, it is not difficult to show that $x^\star$ and $y^\star$ should satisfy $\underline{x} \le x^\star \le \overline{x}$ and $\underline{y} \le y^\star \le \overline{y}$, respectively, where
\begin{align}\label{eqn:over:under:line:xy}
\underline{x} = \min_{k\in\mathcal K} x_k,~\overline{x} = \max_{k\in\mathcal K} x_k,~\underline{y} = \min_{k\in\mathcal K} y_k,~\overline{y} = \max_{k\in\mathcal K} y_k.
\end{align}
This is because if $(x^\star, y^\star)$ lies outside the box specified above, we can always increase the energy transferred to all the $K$ ERs by moving $(x^\star, y^\star)$ into the box. As a result, we can adopt a 2D exhaustive search over the box region $[\underline{x},\overline{x}] \times [\underline{y},\overline{y}]$ to find $(x^\star, y^\star)$.\footnote{Actually, $(x^\star,y^\star)$ should lie within the convex hull of all the ERs' locations $(x_1,y_1), \ldots, (x_K,y_K)$, which is generally a subset of the box $[\underline{x},\overline{x}] \times [\underline{y},\overline{y}]$.} Notice that the optimal solution of $x^\star$ and $y^\star$ to problem \eqref{eqn:x_star:y_star} is generally non-unique.

Given $x^\star$ and $y^\star$, we have the following proposition.
\begin{proposition}\label{proposition:3.1}
The optimal trajectory solution to problem (P1) is given as
\begin{align}\label{eqn:optimal:solution:P1}
x^\star(t) = x^\star,~y^\star(t) = y^\star, \forall t\in\mathcal T.
\end{align}
\end{proposition}
\begin{IEEEproof}
See Appendix \ref{appendix:A}.
\end{IEEEproof}

Proposition \ref{proposition:3.1} indicates that the UAV should hover at one single fixed location $(x^\star,y^\star,H)$ during the whole charging period, referred to as {\it single-location hovering}. Due to the non-uniqueness of the optimal solution $x^\star$ and $y^\star$ to problem \eqref{eqn:x_star:y_star}, such an optimal hovering location $(x^\star,y^\star,H)$ is non-unique in general, as will be shown in an example given in the next subsection for the case of $K=2$. However, this single-location-hovering solution can lead to a severe  ``near-far'' fairness issue in multiuser WPT, as the near ERs in close proximity to the optimal hovering location can receive significantly more energy than the far ERs, especially in a large network with many ERs that are sufficiently separated from each other. To overcome this issue, in Section \ref{sec:max:min} we will consider an alternative problem formulation to maximize the minimum received energy among all ERs to ensure their fairness. Before that, in the following subsection, we derive the optimal solution to problem (P1) in closed-form for the special case of $K=2$ ERs to provide more insights on the sum-energy maximization problem.

\subsection{Special Case with $K=2$ ERs}\label{sec:sum:energy:K2}

In the special case with $K = 2$, without loss of generality we assume $x_1 = -D/2$, $x_2 = D/2$, and $y_1 = y_2 = 0$, where $D$ denotes the distance between the two ERs. Based on Proposition \ref{proposition:3.1}, in this case, the UAV should hover at a fixed location $(x^\star,y^\star,H)$ above the line between the two ERs with $y^\star = 0$. Therefore, it only remains to find $x^\star$.

Towards this end, we first re-express the received power by ER $1$ and ER $2$ as follows, given the UAV's location $(x,0,H)$.
\begin{align}
\hat Q_1(x) = & \frac{\beta_0P}{(x + D/2 )^2 + H^2}, \label{eqn:hat:Q1} \\
\hat Q_2(x) = & \frac{\beta_0P}{(x - D/2 )^2 + H^2}. \label{eqn:hat:Q2}
\end{align}
Accordingly, the function $\psi(x, y)$ in \eqref{eqn:psi} can be simplified as
\begin{align}\label{eqn:hat:psi}
\hat{\psi}(x) \triangleq \sum_{k=1}^2 \hat Q_k(x) =\beta_0P \left(\frac{1}{(x + D/2 )^2 + H^2} + \frac{1}{(x - D/2 )^2 + H^2}\right).
\end{align}
As a result, finding $x^\star$ is equivalent to determining the maximizer of $\hat{\psi}(x)$, i.e., $x^\star = \arg \max_{x} \hat{\psi}(x)$.

\begin{lemma}\label{lemma:3.1}
The function $\hat{\psi}(x)$ has the following properties:
\begin{itemize}
  \item It is symmetric over $x=0$, i.e., $\hat{\psi}( - x) = \hat{\psi}(x), \forall x \in (-\infty,\infty)$.
  \item When $D \le {2H}/{\sqrt{3}}$, $\hat\psi(x)$ is monotonically increasing over $x \in (-\infty,0)$ and decreasing over $x \in (0,\infty)$; hence, $x^\star = 0$ is the unique maximizer of $\hat{\psi}(x)$.
  \item When $D > {2H}/{\sqrt{3}}$, there exists a value
  \begin{align}\label{eqn:xi}
    \xi \triangleq \sqrt{-(D^2/4+H^2) + \sqrt{D^4/4 + H^2D^2}} < D/2,
    \end{align}
    such that $\hat\psi(x)$ is monotonically increasing, decreasing, increasing and decreasing over $x\in  (-\infty,-\xi)$, $(-\xi,0)$, $(0,\xi)$ and $(\xi,\infty)$, respectively; hence, together with the symmetry of the function $\hat{\psi}(x)$ over $x=0$, it follows that $x^\star = -\xi$ and $x^\star = \xi$ are the two equivalent maximizers of $\hat\psi(x)$.
\end{itemize}
\end{lemma}
\begin{IEEEproof}
See Appendix \ref{appendix:B}.
\end{IEEEproof}

Based on Lemma \ref{lemma:3.1}, the optimal solution $x^{\star}$ to maximize $\hat{\psi}(x)$ is found. By using this result together with Proposition \ref{proposition:3.1}, we have the following proposition to solve (P1) for $K = 2$.

\begin{figure}
\centering
 \epsfxsize=1\linewidth
    \includegraphics[width=8cm]{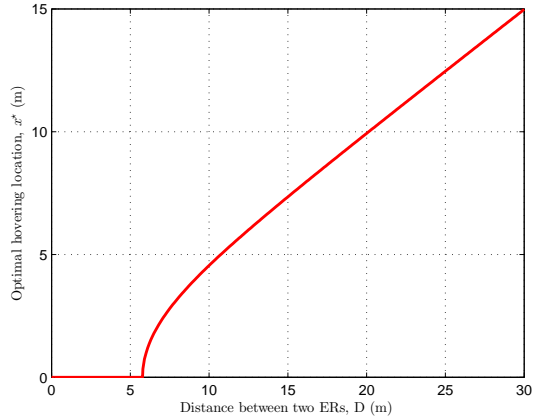}
\caption{The optimal hovering location $x^\star$ versus the distance $D$ between the two ERs in the case of $H = 5$ m.} \label{fig:hovering:location}\vspace{-2em}
\end{figure}

\begin{proposition}\label{proposition:3.2}
In the special case of $K=2$, the optimal solution to (P1) is given as follows.
\begin{itemize}
  \item When $D \le {2H}/{\sqrt{3}}$, we have $x^\star(t) = 0, y^\star(t) = 0, \forall t \in \mathcal T$, i.e., the UAV should hover at the fixed location $(0,0,H)$ above the middle point between the two ERs during the whole charging period.
  \item When $D > {2H}/{\sqrt{3}}$, there are two symmetric optimal solutions to (P1), given by $x^\star(t) = - \xi, y^\star(t) = 0, \forall t \in \mathcal T$, and $x^\star(t) = \xi, y^\star(t) = 0, \forall t \in \mathcal T$, respectively. In other words, the UAV should hover at either $(-\xi,0,H)$ near ER 1 or $(\xi,0,H)$ near ER 2 during the whole charging period.
\end{itemize}
\end{proposition}

For illustration, Fig. \ref{fig:hovering:location} shows the optimal hovering location $x^\star$, with $x^\star(t)=x^\star, \forall t\in \mathcal T$, versus the distance $D$ between the two ERs in the case of $H = 5$ m. It is observed that $x^\star = 0$ when $D\le{2H}/{\sqrt{3}} = 5.77$ m and $0<x^\star<D/2$ (or equivalently $-D/2<-x^\star<0$ for the other symmetric optimal hovering location) when $D>{2H}/{\sqrt{3}}$. It is also observed that as $D$ increases, $x^\star$ becomes closer and eventually converges to $D/2$. These observations are consistent with Proposition \ref{proposition:3.2}. Furthermore, notice that it follows from \eqref{eqn:xi} that $\lim_{D\to\infty} x^\star = \lim_{D\to\infty}\xi = {D}/2$, which can be intuitively explained as follows. When the distance between the two ERs is sufficiently large, the UAV should hover above one of the two ERs to maximize the transferred energy to it, which in turn maximizes the sum-energy transferred to both ERs. In this case, the energy transferred to the other ER becomes negligible. This thus leads to a severe near-far fairness issue between the two ERs.

\section{Min-Energy Maximization Without UAV Speed Constraint}\label{sec:max:min}

In this section, to overcome the aforementioned fairness issue in the sum-energy maximization, we consider an alternative performance metric, namely the min-energy maximization. Specifically, we maximize the minimum received energy among all the $K$ ERs via optimizing the UAV's trajectory $\{x(t),y(t)\}$, subject to the maximum speed constraints in \eqref{eqn:UAV:speed:constraint}. In general, this problem is formulated as
\begin{align}
\text{(P2)}:\max_{\{x(t),y(t)\}}  ~& \min_{k\in\mathcal K} ~E_k(\{x(t),y(t)\}) \nonumber\\
{\mathrm{s.t.}}~~& \eqref{eqn:UAV:speed:constraint}. \nonumber
\end{align}
Problem (P2) is a non-convex optimization problem, and it is more difficult to solve than the sum-energy maximization problem (P1). In particular, the single-location-hovering optimal solution of (P1) given in Proposition \ref{proposition:3.1} is no longer valid for (P2). To tackle (P2), in this section we first consider an ideal case by ignoring the UAV speed constraints in \eqref{eqn:UAV:speed:constraint} and solve the relaxed problem optimally. In practice, the speed constraints in \eqref{eqn:UAV:speed:constraint} can be ignored if the charging duration $T$ and/or the maximum UAV speed $V$ are sufficiently large (see Proposition \ref{proposition:5.1} for a more rigorous argument). For ease of presentation, we rewrite problem (P2) without the constraint \eqref{eqn:UAV:speed:constraint} in the following problem, denoted by (P3).
\begin{align}
\text{(P3)}:\max_{\{x(t),y(t)\}}  ~& \min_{k\in\mathcal K} ~E_k(\{x(t),y(t)\}) \nonumber
\end{align}
In Section \ref{sec:max:min:2}, we will consider the general case of (P2) with the UAV speed constraints \eqref{eqn:UAV:speed:constraint} included, and propose efficient solutions to (P2) based on the optimal solution obtained for (P3).

\subsection{Optimal Solution to Problem (P3)}

Problem (P3) can be equivalently expressed in the following problem by introducing an auxiliary variable $E$.
\begin{align}
\text{(P3.1)}:\max_{\{x(t),y(t)\},E}  ~&  E \nonumber\\
\mathrm{s.t.}~~~~& E_k(\{x(t),y(t)\}) \ge E, \forall k\in\mathcal K.\label{eqn:individual:energy:con}
\end{align}
Though problem (P3.1) is non-convex, it can be shown that it satisfies the so-called time-sharing condition in \cite{YuLui2006}. Therefore, the strong duality holds between (P3.1) and its Lagrange dual problem. As a result, we can optimally solve (P3.1) by using the Lagrange dual method \cite{Boyd:Book}.

Let $\lambda_k \ge 0, k\in\mathcal K$, denote the dual variable associated with the constraint in \eqref{eqn:individual:energy:con} for the $k$th ER. Then the Lagrangian associated with (P3.1) is given as
\begin{align}
&\mathcal L(\{x(t),y(t)\},E,\{\lambda_k\})
=  \big(1-\sum_{k\in\mathcal K} \lambda_k\big)E + \int_{0}^T  \sum_{k\in\mathcal K} \lambda_k Q_k(x(t),y(t))  \text{d}t.
\end{align}
Accordingly, the dual function of (P3.1) is given by
\begin{align}
f(\{\lambda_k\}) = \max_{\{x(t),y(t)\},E} \mathcal L(\{x(t),y(t)\},E,\{\lambda_k\}),\label{eqn:dual:function:f}
\end{align}
for which the following lemma holds.

\begin{lemma}\label{lemma:boundedness}
In order for the dual function $f(\{\lambda_k\})$ to be upper-bounded from above (i.e., $f(\{\lambda_k\}) < \infty$), it must hold that $\sum_{k\in\mathcal K} \lambda_k = 1$.
\end{lemma}
\begin{IEEEproof}
Suppose that $\sum_{k\in\mathcal K} \lambda_k > 1$ (or $\sum_{k\in\mathcal K} \lambda_k < 1$). Then by setting $E\to - \infty$ (or $E\to \infty$), we have $f(\{\lambda_k\}) \to \infty$. Therefore, this lemma is proved.
\end{IEEEproof}

Based on Lemma \ref{lemma:boundedness}, the dual problem of (P3.1) is given by
\begin{align}
\text{(D3.1)}: \min_{\{\lambda_k\}}~ & f(\{\lambda_k\})  \nonumber\\
\mathrm{s.t.}~&\sum_{k\in\mathcal K} \lambda_k = 1 \label{eqn:bound}\\
&\lambda_k \ge 0, \forall k\in\mathcal K. \label{eqn:positive}
\end{align}

Then, we can solve problem (P3.1) by equivalently solving its dual problem (D3.1). Let the feasible set of $\{\lambda_k\}$ characterized by the constraints in \eqref{eqn:bound} and \eqref{eqn:positive} as $\mathcal{X}$. In the following, we first solve problem (\ref{eqn:dual:function:f}) to obtain $f(\{\lambda_k\})$ under any given feasible dual variables $\{\lambda_k\} \in \mathcal X$, then solve (D3.1) to find the optimal $\{\lambda_k\}$ to minimize $f(\{\lambda_k\})$, and finally construct the optimal primal solution to (P3.1).

\subsubsection{Obtaining $f(\{\lambda_k\})$ via Solving Problem \eqref{eqn:dual:function:f} under Given $\{\lambda_k\} \in \mathcal X$}

For any given $\{\lambda_k\} \in \mathcal X$, problem \eqref{eqn:dual:function:f} can be decomposed into the following subproblems.
\begin{align}
\max_{E} ~& (1-\sum_{k\in\mathcal K} \lambda_k)E \label{eqn:subproblem:E}\\
\max_{x(t),y(t)} &\tilde\psi^{\{\lambda_k\}}(x(t),y(t)) \triangleq \sum_{k\in\mathcal K}  \lambda_k Q_k(x(t),y(t))  , ~\forall t\in\mathcal T\label{eqn:subprobelm:q}
\end{align}
In the above, (\ref{eqn:subprobelm:q}) consists of an infinite number of subproblems, each corresponding to a time instant $t$. Let the optimal solutions to \eqref{eqn:subproblem:E} and (\ref{eqn:subprobelm:q}) be denoted by $E^{\{\lambda_k\}}$, as well as $x^{\{\lambda_k\}}(t)$ and $y^{\{\lambda_k\}}(t)$, $\forall t\in\mathcal T$, respectively.

As for subproblem (\ref{eqn:subproblem:E}), since $1-\sum_{k\in\mathcal K} \lambda_k =0$ holds for any given $\{\lambda_k\}\in \mathcal X$, its objective value is always zero. In this case, we can choose any arbitrary real number as the optimal solution $E^{(\{\lambda_k\})}$ for the purpose of obtaining the dual function $f(\{\lambda_k\})$.

On the other hand, note that the subproblems in (\ref{eqn:subprobelm:q}) are identical for all time instant $t\in \mathcal T$. Therefore, we can drop the time index $t$ and re-express each problem in \eqref{eqn:subprobelm:q} as
\begin{align}\label{eqn:bar:psi}
\max_{x,y}~ & \tilde\psi^{\{\lambda_k\}}(x,y).
\end{align}
Similarly as for problem \eqref{eqn:x_star:y_star}, problem \eqref{eqn:bar:psi} has two optimization variables, and the optimal solution of $x^{\{\lambda_k\}}$ and $y^{\{\lambda_k\}}$ satisfies $\underline{x} \le x^{\{\lambda_k\}} \le \overline{x}$ and $\underline{y} \le y^{\{\lambda_k\}} \le \overline{y}$, with $\underline{x}$, $\overline{x}$, $\underline{y}$, and $\overline{y}$ given in \eqref{eqn:over:under:line:xy}. As a result, we can adopt a 2D exhaustive search over the box region $[\underline{x},\overline{x}] \times [\underline{y},\overline{y}]$ to find the optimal $x^{\{\lambda_k\}}$ and $y^{\{\lambda_k\}}$. Accordingly, the optimal solution to problem \eqref{eqn:subprobelm:q} is given by
\begin{align}
x^{\{\lambda_k\}}(t) = x^{\{\lambda_k\}},~y^{\{\lambda_k\}}(t) = y^{\{\lambda_k\}}, \forall t\in\mathcal T.
\end{align}
Note that the optimal solution of $x^{\{\lambda_k\}}$ and $y^{\{\lambda_k\}}$ to \eqref{eqn:bar:psi} is generally non-unique, and we can arbitrarily choose any one of them for obtaining the dual function $f(\{\lambda_k\})$.\footnote{When the optimal solution $x^{\{\lambda_k\}}(t)$'s, $y^{\{\lambda_k\}}(t)$'s, and $E^{(\{\lambda_k\})}$ are not unique, they may not be optimal for the primal problem (P3.1) after the dual problem is solved. As a result, an additional step is required to obtain the optimal primal solution of $E$, $x(t)$'s, and $y(t)$'s to (P3.1), as will be shown later in Section \ref{sec:construct}.} By substituting $E^{\{\lambda_k\}}$, $x^{\{\lambda_k\}}(t)$'s and $y^{\{\lambda_k\}}(t)$'s into problem (\ref{eqn:dual:function:f}), the dual function $f(\{\lambda_k\})$ is obtained.

\subsubsection{Finding Optimal Dual Solution to (D3.1)}

With $f(\{\lambda_k\})$ obtained, we then solve the dual problem (D3.1) to find the optimal $\{\lambda_k\}$ to minimize $f(\{\lambda_k\})$. Note that the dual function $f(\{\lambda_k\})$ is always convex but generally non-differentiable \cite{Boyd:Book}. As a result, problem (D3.1) can be solved by subgradient-based methods such as the ellipsoid method \cite{BoydII}. Note that the subgradient of the objective function $f(\{\lambda_k\})$ is given by \begin{align*}
\mv{s}_0(\lambda_1,\ldots,\lambda_K) = \left[ T Q_1(x^{(\{\lambda_k\})},y^{(\{\lambda_k\})}),\ldots, T Q_K(x^{(\{\lambda_k\})},y^{(\{\lambda_k\})})\right],
\end{align*}
where $E^{(\{\lambda_k\})} = 0$ is chosen for simplicity. Furthermore, the equality constraint in (\ref{eqn:bound}) can be viewed as two inequality constraints $1 - \sum_{k\in\mathcal K} \lambda_k \le 0$ and $-1 + \sum_{k\in\mathcal K} \lambda_k \le 0$, whose subgradients are given by $\mv{s}_1(\lambda_1,\ldots,\lambda_K) = -\mv{e}$ and $\mv{s}_2(\lambda_1,\ldots,\lambda_K) = \mv{e}$, respectively, where $\mv e$ denotes an all-one vector. We denote the obtained optimal dual solution to (D3.1) as $\{\lambda_k^*\}$.


\subsubsection{Constructing Optimal Primal Solution to (P3.1)}\label{sec:construct}

Based on the optimal dual solution $\{\lambda_k^*\}$ to (D3.1), we need to obtain the optimal primal solution to (P3.1), denoted by $\{x^*(t)\}$, $\{y^*(t)\}$, and $E^*$. It is worth noting that when using the Lagrange dual method to solve problem (P3.1) via the dual problem (D3.1), the optimal solution to problem (\ref{eqn:dual:function:f}) under the optimal dual solution $\{\lambda_k^*\}$ (i.e., $x^{\{\lambda_k^*\}}(t)$'s, $y^{\{\lambda_k^*\}}(t)$'s, and $E^{\{\lambda_k^*\}}$) is the optimal primal solution to (P3.1), if such a solution is {\it unique and primal feasible} \cite{Boyd:Book}. On the other hand, when the optimal $x^{\{\lambda_k^*\}}(t)$'s, $y^{\{\lambda_k^*\}}(t)$'s, and $E^{\{\lambda_k^*\}}$ to problem (\ref{eqn:dual:function:f}) are non-unique, they may not be feasible nor optimal to problem (P3.1) in general. In the latter case, we need to time-share among these non-unique optimal solutions as follows to construct the optimal primal solution $\{x^*(t)\}$, $\{y^*(t)\}$, and $E^*$ to (P3.1).

With the optimal dual solution $\{\lambda_k^*\}$, suppose that problem \eqref{eqn:bar:psi} has a total number of $\Gamma \ge 1$ optimal location solutions to maximize $\psi^{\{\lambda^*_k\}}(x,y)$, denoted by $(x^*_1,y^*_1), \ldots, (x^*_\Gamma,y^*_\Gamma)$. Let $Q_k(x^*_\gamma,y^*_\gamma)$ denote the corresponding received power at each ER $k\in\mathcal K$ when the UAV stays at the location  $(x^*_\gamma,y^*_\gamma,H)$. In this case, proper time-sharing among the $\Gamma$ solutions is necessary for constructing the optimal primal solution to (P3.1). Time-sharing means that the UAV should hover at each of these different locations for a certain portion of the total duration $T$. Let $\tau_\gamma$ denote the optimal hovering duration at $(x^*_\gamma,y^*_\gamma,H)$. Then, the optimal $\tau_\gamma^*$'s, together with the maximum min-energy $E^*$, can be obtained by solving the following problem.
\begin{align}
\max_{\{\tau_\gamma \ge 0\},E}~&E \nonumber\\
\mathrm{s.t.}~~& \sum_{\gamma = 1}^{\Gamma} \tau_\gamma Q_k(x^*_\gamma,y^*_\gamma) \ge E, \forall k\in\mathcal K \nonumber\\
& \sum_{\gamma = 1}^{\Gamma} \tau_\gamma = T. \label{eqn:problem:time:sharing}
\end{align}
Note that problem \eqref{eqn:problem:time:sharing} is a linear program (LP), which can be solved by using standard convex optimization techniques \cite{Boyd:Book}. As a result, the optimal solution of $E^*$ to (P3.1) is found. Finally, we obtain the optimal trajectory solution of $\{x^*(t),y^*(t)\}$ to (P3.1) (and thus (P3)), which is given in the following proposition based on the above time-sharing property, with the proof omitted for brevity.
\begin{proposition}\label{proposition:opt:P5.1}
Partition the whole charging period into $\Gamma$ portions, denoted by $\mathcal T_1, \ldots, \mathcal T_\Gamma$, where $\mathcal T_\gamma = (\sum_{i = 1}^{\gamma-1}\tau^*_i,\sum_{i = 1}^{\gamma}\tau^*_i]$ with duration $\tau^*_\gamma$ for $\gamma \geq 1$. Then, the optimal trajectory solution of $\{x^*(t),y^*(t)\}$ to (P3.1) or (P3) is given by
\begin{align}\label{eqn:optimal:trajectory:P3}
x^*(t) = x^*_\gamma, y^*(t) = y^*_\gamma, \forall t\in \mathcal T_\gamma, \gamma\in\{1,\ldots,\Gamma\},
\end{align}
where $\mathcal T_\gamma \cap \mathcal T_\zeta = \phi, \forall \gamma\neq \zeta$, and $\bigcup_{\gamma=1}^{\Gamma}\mathcal T_\gamma = \mathcal T$.
\end{proposition}

In summary, we present the algorithm for solving (P3.1) or (P3) as Algorithm 1 in Table \ref{Table:I}.

\begin{table}[!t]\scriptsize
\caption{Algorithm 1: Algorithm for Solving Problem (P3.1) or (P3)} \centering
\begin{tabular}{|p{15cm}|}
\hline\vspace{0.01cm}
\begin{itemize}
\item[a)] {\bf Initialization:} Given an ellipsoid ${\cal E}(\mv\lambda,{\mv A})$ containing $\mv\lambda^* = [\lambda_1^*,\ldots,\lambda_K^*]$, where $\mv \lambda =[\lambda_1,\ldots,\lambda_K]$ is the center point of ${\cal E}$ and the positive definite matrix ${\mv A}$ characterizes the size of ${\cal E}$.
\item[b)] {\bf Repeat:}
    \begin{itemize}
    \item[1)] Obtain $x^{\{\lambda_k\}}$ and $y^{\{\lambda_k\}}$ to maximize $\psi^{\{\lambda_k\}}(x,y)$ via a 2D exhaustive search over the region $[\underline{x},\overline{x}] \times [\underline{y},\overline{y}]$;
    \item[2)] Update $\{\lambda_k\}$ using the ellipsoid method by noting that the subgradient of $f(\{\lambda_k\})$ is $\mv{s}_0(\lambda_1,\ldots,\lambda_K)$ \cite{BoydII}.
    \end{itemize}
\item[c)] {\bf Until} $\{\lambda_k\}$ converge with a prescribed accuracy.
\item[d)] {\bf Set} $\lambda_k^* \gets \lambda_k, \forall k\in\mathcal K$.
\item[e)] {\bf Output}: Obtain all the $\Gamma$ optimal locations to maximize the function $\psi^{\{\lambda^*_k\}}(x,y)$ as $(x^*_1,y^*_1), \ldots, (x^*_\Gamma,y^*_\Gamma)$; then solve the LP in (\ref{eqn:problem:time:sharing}) to obtain $\{\tau_\gamma^*\}_{\gamma=1}^{\Gamma}$ and $E^*$; and finally obtain the optimal trajectory solution of $\{x^*(t),y^*(t)\}$ to problem (P3.1) or (P3) as in \eqref{eqn:optimal:trajectory:P3}.
\end{itemize}\\ \hline
\end{tabular}\label{Table:I}\vspace{-1em}
\end{table}

\begin{remark}
Note that Proposition \ref{proposition:opt:P5.1} implies that to maximize the min-energy transferred to the $K$ ERs, the UAV should hover above a number of fixed locations during the charging period, and the optimal hovering locations (i.e., $x^*_\gamma$'s and $y^*_\gamma$'s) are generally different from the locations of the ERs (i.e., $x_k$'s and $y_k$'s). We refer to such a design as {\it multi-location hovering}. Different from the single-location hovering for sum-energy maximization, the result here shows that the UAV should in general hover over different locations so as to balance the energy transferred to all ERs.
\end{remark}

\subsection{Special Case with $K=2$ ERs}\label{sec:opt:solution:P3:K2}

To provide more insights, in this subsection we consider the min-energy maximization in the special case with $K=2$ ERs, by assuming $x_1 = -D/2, x_2 = D/2$, and $y_1=y_2=0$, similarly as in Section \ref{sec:sum:energy:K2}. In this case, by setting $y(t) = 0, \forall t\in \mathcal K$, problem (P3) can be simplified as
\begin{align}
\text{(P3-2ER)}:\max_{\{{x}(t)\}}  ~& \min(\hat E_1(\{x(t)\}),\hat E_2(\{x(t)\})).
\end{align}
\begin{proposition}\label{proposition:4.1}
The optimal solution to problem (P3-2ER) is given as follows by considering two different cases:
\begin{itemize}
  \item If $D \le 2H/\sqrt{3}$, then the UAV should hover at the fixed location $(0,0,H)$ during the whole charging period, i.e., $x^*(t) = 0, \forall t \in \mathcal T$.
  \item If $D > 2H/\sqrt{3}$, then the UAV should hover over the two symmetric locations $(-\xi,0,H)$ and $(\xi,0,H)$ with equal durations, e.g., $x^*(t) = -\xi, \forall t \in [0,T/2)$, and $x^*(t) = \xi, \forall t \in [T/2,T]$, with $\xi$ given in \eqref{eqn:xi}.
\end{itemize}
\end{proposition}
\begin{IEEEproof}
See Appendix \ref{appendix:C:new}.
\end{IEEEproof}

By comparing Proposition \ref{proposition:4.1} versus Proposition \ref{proposition:3.2}, it is observed that in the case with $K=2$ ERs, each of the optimal hovering locations for the min-energy maximization is also optimal for the corresponding case in the sum-energy maximization. Thus, the achieved sum-energy in problem (P3-2ER) is identical to that in (P1) when $K=2$. This implies that without the maximum UAV speed constraint, the multi-location-hovering solution to (P3-2ER) can maximize the sum-energy received by the two ERs, while ensuring their energy fairness thanks to the time sharing. Nevertheless, it is worth noting that in the general case with $K>2$, the optimal hovering locations for (P3) are generally different from that for (P1). This is due to the fact that $x^\star$ and $y^\star$ are the maximizer of the sum-power $\psi(x,y)$, but $x^*_\gamma$'s and $y^*_\gamma$'s are the maximizers of the weighted sum-power $\tilde\psi^{\{\lambda_k\}}(x(t),y(t))$, which is different from $\psi(x,y)$ if any two $\lambda_{k}$'s are not identical in the case of $K>2$.

\section{Min-Energy Maximization With UAV Speed Constraint}\label{sec:max:min:2}

In this section, we consider the general min-energy maximization problem (P2) by including the practical UAV maximum speed constraints in \eqref{eqn:UAV:speed:constraint}. This problem is difficult to be solved globally optimally in general with $K>2$. To tackle this problem, we propose two suboptimal solutions inspired by the optimal solution obtained previously for problem (P3) in the ideal case without the UAV maximum speed constraint.

\subsection{Successive Hover-and-Fly Trajectory Design for Problem (P2)}

In this subsection, we propose a {\it successive hover-and-fly} trajectory design to solve problem (P2) based on the optimal solution obtained for (P3) in the ideal case. Recall that the optimal solution to (P3) corresponds to $\Gamma$ optimal hovering locations, i.e., $\{(x^*_\gamma,y^*_\gamma,H)\}_{\gamma=1}^\Gamma$. In the proposed trajectory design with the maximum speed constraint, the UAV sequentially hovers at each of these locations for a certain duration and flies from one location to another with the maximum speed $V$. As a result, to find the optimal successive hover-and-fly trajectory, we need to first determine the UAV's traveling path to visit all the $\Gamma$ locations with the minimum flying distance so as to minimize the total flying time, and then optimize the hovering time at each of these locations for the remaining time in the charging duration.

\subsubsection{Flying Distance Minimization to Visit $\Gamma$ Hovering Locations}

First, we determine the UAV's traveling path to visit all the $\Gamma$ hovering locations with the minimum flying distance. For ease of description, let $d_{\gamma,\zeta} \triangleq \sqrt{(x_\gamma^* - x_\zeta^*)^2+(y_\gamma^* - y_\zeta^*)^2}$ denote the distance between the $\gamma$th hovering location $(x_\gamma^*,y^*_\gamma,H)$ and the $\zeta$th hovering location $(x_\zeta^*,y^*_\zeta,H)$. We define a binary variable $f_{\gamma,\zeta}$ for any $\gamma,\zeta \in \{1,\ldots,\Gamma\}, \gamma\neq \zeta$, where $f_{\gamma,\zeta} = 1$ indicates that the UAV should fly from the $\gamma$th hovering location $(x_\gamma^*,y^*_\gamma,H)$ to the $\zeta$th hovering location $(x_\zeta^*,y^*_\zeta,H)$, and $f_{\gamma,\zeta} = 0$ otherwise. The trajectory design problem thus becomes determining $\{f_{\gamma,\zeta}\}$ to minimize $\sum_{\gamma=1}^{\Gamma} \sum_{\zeta=1,\zeta\neq \gamma}^{\Gamma}f_{\gamma,\zeta}d_{\gamma,\zeta}$, provided that each of the $\Gamma$ locations is visited exactly once.

The flying distance minimization problem considered here is reminiscent of the well-known traveling salesman problem (TSP) (see, e.g., \cite{TSP1,TSP2,TSP3}), with the following difference. In the standard TSP, the salesman (or equivalently the UAV of our interest) needs to return to the origin city (the initial hovering location) after visiting all these cities (or hovering locations here); but our flying distance minimization problem does not have such a requirement since the initial and final hovering locations can be optimized. Fortunately, it has been shown in \cite{TSPw_o_origin_end} that our flying distance minimization problem can be transformed to the standard TSP as follows. First, we add a dummy hovering location, namely the $(\Gamma+1)$-th hovering location, whose distances to all the existing $\Gamma$ hovering locations are $0$, i.e., $d_{\Gamma+1,\gamma} = d_{\gamma,\Gamma+1} = 0, \forall \gamma \in \{1,\ldots,\Gamma\}$. Note that this dummy hovering location is a virtual node that does not exist physically. Then, we obtain the desirable traveling path by solving the standard TSP problem for the $\Gamma+1$ hovering locations,\footnote{Note that although the TSP is an NP-hard problem in combinatorial optimization, various heuristic and approximation algorithms have been proposed to give efficient high-quality solutions for it (see, e.g., \cite{TSP1,TSP2,TSP3}). In particular, it has been shown in \cite{TSP3} that the TSP problem can be formulated as a binary integer program, by incorporating a set of constraints to ensure there is only a single tour connecting all visited locations. The binary integer program can be solved via CVX \cite{CVX} by using the Mosek solver that supports the integer program (see \url{https://mosek.com/} for details).} and then removing the two edges associated with the dummy location. For the obtained traveling path, we define the permutation $\pi(\cdot)$ over the set $\{1,\ldots,\Gamma\}$, such that the UAV first visits the $\pi(1)$-th hovering location, followed by the $\pi(2)$-th, the $\pi(3)$-th, until the $\pi(\Gamma)$-th hovering location at last. In this case, the resulting flying distance and flying duration  with the maximum speed $V$ are given as $D_{\text{fly}} = \sum_{\gamma=1}^{\Gamma-1} d_{\pi(\gamma),\pi(\gamma+1)}$, and $T_{\text{fly}} = D_{\text{fly}}/V$, respectively. We denote the corresponding trajectory as $\{\hat{x}(t),\hat{y}(t)\}_{t=0}^{T_{\text{fly}}}$.

It is worth noting that the above traveling path is only feasible when the charging duration $T$ is no smaller than $T_{\text{fly}}$, i.e., $T \ge T_{\text{fly}}$, since otherwise the charging duration is not sufficient for the UAV to visit all the $\Gamma$ hovering locations. In the following, we first determine the hovering time allocation over different locations in the case with $T \ge T_{\text{fly}}$, and then refine the trajectory design in the case with $T < T_{\text{fly}}$.

\subsubsection{Hovering Time Allocation When $T \ge T_{\text{fly}}$}\label{sec:hovering:time:allocation}

First, we consider the case when $T \ge T_{\text{fly}}$. With the above traveling path $\{\hat x(t), \hat y(t)\}_{t=0}^{T_{\text{fly}}}$, the trajectory design problem remains to allocate the hovering duration $T - T_{\text{fly}}$ among the $\Gamma$ locations to maximize the min-energy transferred to all the $K$ ERs. Note that based on the traveling path $\{\hat{x}(t),\hat{y}(t)\}_{t=0}^{T_{\text{fly}}}$, we can obtain the received energy by each ER $k\in\mathcal K$ during the UAV's flying time as  $E_k^{\text{fly}} = \int_{0}^{T_{\text{fly}}}Q_k(\hat{x}(t),\hat{y}(t))\mathrm{d}t$, with $Q_k(\cdot,\cdot)$ given in \eqref{eqn:harvested:power}. Also, recall that $Q_k(x^*_\gamma,y^*_\gamma)$ denotes the received power at ER $k\in\mathcal K$ when the UAV hovers at the location  $(x^*_\gamma,y^*_\gamma,H)$. Then the optimal hovering durations, denoted as $\tau_\gamma^{**}$'s, together with the corresponding maximum min-energy of the $K$ ERs, denoted by $E^{**}$, can be obtained by solving the following LP.
\begin{align}
\max_{\{\tau_\gamma \ge 0\},E}~&E \nonumber\\
\mathrm{s.t.}~~& \sum_{\gamma = 1}^{\Gamma} \tau_\gamma Q_k(x^*_\gamma,y^*_\gamma) + E_k^{\text{fly}} \ge E, \forall k\in\mathcal K \nonumber\\
& \sum_{\gamma = 1}^{\Gamma} \tau_\gamma = T -  T_{\text{fly}}. \label{eqn:problem:time:sharing:2}
\end{align}

With the optimal permutation $\pi(\cdot)$ and the optimal hovering durations $\{\tau_\gamma^{**}\}$ obtained, the successive hover-and-fly trajectory is finalized, which can be summarized as follows. By dividing the charging period into $2\Gamma - 1$ slots; in the $(2\gamma-1)$-th slot with duration $\tau_{\pi(\gamma)}^{**}$, $\gamma \in \{1, \ldots, \Gamma\}$, the UAV hovers at the $\pi(\gamma)$-th hovering location $(x^*_{\pi(\gamma)},y^*_{\pi(\gamma)},H)$; and in the $(2\gamma)$-th slot, $\gamma \in \{1, \ldots, \Gamma-1\}$, the UAV flies from the $\pi(\gamma)$-th hovering location $(x^*_{\pi(\gamma)},y^*_{\pi(\gamma)},H)$ to the $\pi(\gamma+1)$-th hovering location $(x^*_{\pi(\gamma+1)},y^*_{\pi(\gamma+1)},H)$ with its maximum speed $V$.

\begin{proposition}\label{proposition:5.1}
When the charging duration $T$ is sufficiently large such that $T\gg T_{\text{fly}}$, the successive hover-and-fly trajectory design is asymptotically optimal for problem (P2).
\end{proposition}
\begin{IEEEproof}
When $T\gg T_{\text{fly}}$, the flying time is negligible and thus the successive hover-and-fly trajectory is equivalent to the optimal multi-location-hovering solution to (P3). In this case, the objective value achieved by the successive hover-and-fly trajectory for (P2) is asymptotically approaching the optimal value of (P3), which actually serves as the upper bound for that of (P2). Therefore, the proposed trajectory design is asymptotically optimal for (P2) when $T\gg T_{\text{fly}}$.
\end{IEEEproof}

\subsubsection{Trajectory Redesign When $T < T_{\text{fly}}$}

In this subsection, we consider the case when $T < T_{\text{fly}}$. In this case, the UAV traveling path $\{\hat x(t), \hat y(t)\}_{t=0}^{T_{\text{fly}}}$ based on the TSP solution is no longer feasible since the charging time is not sufficient for the UAV to visit all the $\Gamma$ hovering locations. To address this issue, we first find the solution to (P2) when $T$ is sufficiently small (i.e., $T \to 0$) such that the UAV can only hover at one single location, and then reconstruct a modified successive hover-and-fly trajectory for the case of $T < T_{\text{fly}}$.

First, when $T \to 0$, the UAV should hover at one single fixed location, denoted by $(x_{\text{fix}},y_{\text{fix}},H)$, where $x_{\text{fix}}$ and $y_{\text{fix}}$ can be obtained by solving the following problem via a 2D exhaustive search over $(\underline{x},\overline{x}) \times (\underline{y},\overline{y})$.
\begin{align}\label{eqn:fix}
(x_{\text{fix}},y_{\text{fix}}) = \arg\max_{x,y}  ~& \min_{k\in\mathcal K} ~Q_k(x,y).
\end{align}

Next, we reconstruct the trajectory as follows by down-scaling the previously obtained traveling path $\{(\hat{x}(t),\hat{y}(t),H)\}_{t=0}^{T_{\text{fly}}}$ for the case of $T=T_{\text{fly}}$ linearly towards the center point $(x_{\text{fix}},y_{\text{fix}},H)$, such that the resulting total flying distance equals $VT$.
\begin{align}\label{eqn:scaling}
x^{**}(t) = \hat{x}(t/\kappa) + (1-\kappa) (x_{\text{fix}} - \hat{x}(t/\kappa)),~y^{**}(t) = \hat{y}(t/\kappa) + (1-\kappa) (y_{\text{fix}} - \hat{y}(t/\kappa)),~\forall t\in [0,T],
\end{align}
where $\kappa = T/T_{\text{fly}} < 1$ denotes the linear scaling factor. Note that when $T\to 0$, we have $\kappa \to 0$, and the above redesigned trajectory reduces to hovering at one single fixed location $(x_{\text{fix}},y_{\text{fix}},H)$; while $T\to T_{\text{fly}}$, we have $\kappa \to 1$, and the above redesigned trajectory becomes identical to the TSP-based trajectory $\{(\hat{x}(t),\hat{y}(t),H)\}_{t=0}^{T_{\text{fly}}}$.

%

We summarize the overall algorithm for the proposed successive hover-and-fly trajectory design as Algorithm 2 in Table \ref{table:II}, for both the cases of $T \ge T_{\text{fly}}$ and $T < T_{\text{fly}}$.

\begin{table}[!t]\scriptsize
\caption{Algorithm 2: Successive Hover-and-Fly Trajectory Design for Solving Problem (P2)} \centering
\begin{tabular}{|p{15cm}|}
\hline\vspace{0.01cm}
\begin{itemize}
\item[a)] Solve problem (P3.1) by Algorithm 1 in Table \ref{Table:I} to find the $\Gamma$ hovering locations $\{(x_{\gamma}^*,y^*_{\gamma},H)\}_{\gamma=1}^\Gamma$.
\item[b)] Add a dummy hovering location, namely the $(\Gamma+1)$-th hovering location, and set its distances to all the existing $\Gamma$ hovering locations as $0$.
\item[c)] Obtain the desirable traveling path $\{\hat{x}(t),\hat{y}(t)\}_{t=0}^{T_{\text{fly}}}$ by solving the standard TSP problem for the $\Gamma+1$ hovering locations and then removing the two edges associated with the dummy location, where $T_{\text{fly}}$ denotes the total flying time.
\item[d)] If $T \ge T_{\text{fly}}$, then find the optimal hovering time allocations $\tau_\gamma^{**}$'s by solving problem \eqref{eqn:problem:time:sharing:2}; accordingly, obtain the corresponding trajectory $\{x^{**}(t),y^{**}(t)\}$ as in Section \ref{sec:hovering:time:allocation}.
\item[e)] Otherwise, if $T < T_{\text{fly}}$, then obtain the trajectory $\{x^{**}(t),y^{**}(t)\}$ based on \eqref{eqn:scaling}.
\end{itemize}\\ \hline
\end{tabular}\label{table:II}\vspace{-2em}
\end{table}

\subsubsection{Optimality in the Case With $K=2$ ERs}

In the special two-ER case with $x_1 = -D/2, x_2 = D/2$, and $y_1=y_2=0$, the min-energy maximization problem with the maximum speed constraint is simplified as follows by setting $y(t) = 0, \forall t\in\mathcal T$.
\begin{align}
\text{(P2-2ER)}:\max_{\{{x}(t)\}}  ~& \min(\hat E_1(\{x(t)\}),\hat E_2(\{x(t)\}))\nonumber\\
\mathrm{s.t.}~&|\dot{x}(t)| \le V, \forall t\in\mathcal T. \nonumber
\end{align}
As it has been shown in Proposition \ref{proposition:4.1} that for problem (P3-2ER) without the maximum speed constraint, if $D \le 2H/\sqrt{3}$, then there is one optimal hovering location $(0,0,H)$; while if $D > 2H/\sqrt{3}$, then there are two symmetric optimal hovering locations $(-\xi,0,H)$ and $(\xi,0,H)$ with equal time allocation. By applying this result in Algorithm 2, the corresponding successive hover-and-fly trajectory for (P2-2ER) is obtained as follows by considering three different cases:
\begin{itemize}
  \item If $D \le 2H/\sqrt{3}$, then the UAV should hover at the fixed location $(0,0,H)$ above the middle point between the two ERs during the whole charging period, i.e., $x^*(t) = 0, \forall t \in \mathcal T$.
  \item If $D > 2H/\sqrt{3}$ and $T \le 2\xi/V = T_{\text{fly}}$ with $\xi$ given in \eqref{eqn:xi}, then the UAV should fly from the location $(-VT/2,0,H)$ to $(VT/2,0,H)$ with the maximum speed $V$, i.e., $x^{**}(t) = -VT/2 + Vt, \forall t \in \mathcal T$.
  \item If $D > 2H/\sqrt{3}$ and $T > 2\xi/V$, then the UAV should follow a successive hover-and-fly trajectory: first, the UAV hovers at the location $(-\xi,0,H)$ for the duration $t \in  [0,T/2-\xi/V]$; next, it flies from $(-\xi,0,H)$ to $(\xi,0,H)$ with the maximum speed $V$ during the time interval $t \in (T/2-\xi/V,T/2+\xi/V)$; finally, the UAV hovers at the location $(\xi,0,H)$ for the remaining time $t \in [T/2+\xi/V,T]$. In other words, the optimal UAV trajectory is
\begin{align}\label{eqn:hovering:flying:hovering}
x^{**}(t) =
\left\{
\begin{array}{ll}
-\xi, &t \in [0,T/2-\xi/V]\\
Vt -VT/2,& t \in  (T/2-\xi/V,T/2+\xi/V)\\
\xi,&t \in [T/2+\xi/V,T].
\end{array}
\right.
\end{align}
\end{itemize}

\begin{proposition}\label{proposition:4}
The above successive hover-and-fly trajectory solution is optimal for problem (P2-2ER) in the case of $K=2$ ERs .
\end{proposition}
\begin{IEEEproof}
See Appendix \ref{appendix:D}.
\end{IEEEproof}

\begin{remark}
Proposition \ref{proposition:4} provides important insights on how to maximize the minimum or equal energy transferred to the two ERs. First, when the two ERs are close to each other with $D \le 2H/\sqrt{3}$, the UAV hovers at one fixed location $(0,0,H)$ during the whole charging period, and this solution is also optimal for problem (P1) to maximize the sum-energy of the two ERs (see Proposition \ref{proposition:3.2}). Next, consider the case when the two ERs are located farther apart with $D > 2H/\sqrt{3}$. If the charging duration is short (i.e., $T \le 2\xi/V$), then the UAV should keep flying at its maximum speed from one ER to the other by following a symmetric trajectory around the middle point $(0,0, H)$, without hovering over any of them due to the insufficient charging time. On the other hand, if the charging duration is sufficiently long (i.e., $T > 2\xi/V$), then the UAV should hover at the two symmetric locations $(-\xi,0,H)$ and $(\xi,0,H)$ with equal time and travel from one location to the other with the maximum speed. The optimal trajectory in this case is different from that for sum-energy maximization in Proposition \ref{proposition:3.2}, where the UAV should hover at only one fixed location $(-\xi,0,H)$ or $(\xi,0,H)$ for all the time. Therefore, the fairness is achieved by the UAV traveling between different hovering locations, though a certain loss in the total energy transferred to the two ERs is incurred.
\end{remark}

\subsection{SCP-Based Trajectory Design for Problem (P2)}

In this subsection, we develop an alternative SCP-based algorithm to find a locally optimal solution to problem (P2). Note that the SCP-based trajectory design has been studied for throughput or energy efficiency maximization for UAV-enabled wireless communication systems \cite{ZengZhangLim2016b,Zeng17}, but the results cannot be directly applied for UAV-enabled WPT systems considered in this paper. With the SCP-based trajectory design, we first discretize the whole charging duration into a finite number of $N$ time slots, each with duration $\Delta = T/N$. Note that the duration $\Delta$ is chosen to be sufficiently small, such that we can assume that the UAV location is approximately unchanged during each slot $n$, which is denoted as $(x[n],y[n],H)$, $n \in \mathcal N \triangleq \{1,\ldots,N\}$. In this case, the received energy by each ER $k\in\mathcal K$ at slot $n$ is given by
\begin{align}
\hat{E}_k(x[n],y[n]) =  \frac{\beta_0P \Delta}{(x[n] - x_k)^2 + (y[n] - y_k)^2 + H^2}.
\end{align}
Accordingly, the min-energy maximization problem (P2) over the continuous trajectory $\{x(t),y(t)\}$ can be reformulated as follows over the discretized trajectory variables $\{x[n],y[n]\}_{n=1}^N$.
\begin{align}
&\max_{\{x[n],y[n]\}}   \min_{k\in\mathcal K} \sum_{n=1}^{N} \hat{E}_k(x[n],y[n]) \label{eqn:speed:discretized:P3}\\
&{\mathrm{s.t.}}~(x[n] - x[n-1])^2 + (y[n] - y[n-1])^2 \le V^2\Delta^2, \forall n\in\{2,\ldots,N\},\label{eqn:speed:discretized:P3:constraint}
\end{align}
where the constraints in \eqref{eqn:speed:discretized:P3:constraint} correspond to the discretized version of the maximum speed constraints in \eqref{eqn:UAV:speed:constraint}. Note that the constraints in \eqref{eqn:speed:discretized:P3:constraint}  are all convex but the objective function in \eqref{eqn:speed:discretized:P3} is not concave. Therefore, problem \eqref{eqn:speed:discretized:P3} is a non-convex optimization problem.

For the non-convex optimization problem  \eqref{eqn:speed:discretized:P3}, we obtain a locally optimal solution by proposing an SCP-based algorithm, which is operated in an iterative manner to successively maximize a lower bound of the objective function in \eqref{eqn:speed:discretized:P3} at each iteration. Particularly, let $\{x^{(0)}[n],y^{(0)}[n]\}$ denote the initial trajectory and $\{x^{(i)}[n],y^{(i)}[n]\}$ the obtained trajectory after iteration $i \ge 1$. We have the following lemma.
\begin{lemma}\label{lemma:4.3}
For any given $\{x^{(i)}[n],y^{(i)}[n]\}$, $i\ge 0$, it follows that
\begin{align}\label{eqn:lower:bound}
\hat{E}_k(x[n],y[n]) \ge & \hat{E}^{(i)}_k(x[n],y[n]),\forall k\in\mathcal K,n \in \mathcal N,
\end{align}
where
\begin{align}
\hat{E}^{(i)}_k&(x[n],y[n]) \triangleq  \frac{2\beta_0P \Delta}{(x^{(i)}[n] - x_k)^2 + (y^{(i)}[n] - y_k)^2 + H^2} \nonumber\\
&- \frac{\beta_0P \Delta({(x[n] - x_k)^2 + (y[n] - y_k)^2 + H^2})}{({(x^{(i)}[n] - x_k)^2 + (y^{(i)}[n] - y_k)^2 + H^2})^2}.\label{eqn:hat:E}
\end{align}
The inequalities in \eqref{eqn:lower:bound} are tight for $x[n] = x^{(i)}[n]$ and $y[n]=y^{(i)}[n]$, i.e.,
\begin{align}\label{eqn:strict:eqality}
\hat{E}_k(x^{(i)}[n],y^{(i)}[n]) = & \hat{E}^{(i)}_k(x^{(i)}[n],y^{(i)}[n]),\forall k\in\mathcal K,n \in \mathcal N.
\end{align}
\end{lemma}
\begin{IEEEproof}
See Appendix \ref{appendix:E}.
\end{IEEEproof}

Based on Lemma \ref{lemma:4.3}, at each iteration $i + 1$, we optimize over $\{x[n],y[n]\}$ by replacing $\hat{E}_k(x[n],y[n])$'s in problem \eqref{eqn:speed:discretized:P3} with their respective lower bounds $\hat{E}^{(i)}_k(x[n],y[n])$ in (\ref{eqn:hat:E}). More specifically, the discretized trajectory is updated as
\begin{align}
&\{x^{(i+1)}[n],y^{(i+1)}[n]\} = \arg\max_{\{x[n],y[n]\}} ~ \min_{k\in\mathcal K} \sum_{n=1}^{N} \hat{E}^{(i)}_k(x[n],y[n]),~
{\mathrm{s.t.}}~(\ref{eqn:speed:discretized:P3:constraint}).  \label{eqn:iteration}
\end{align}
Note that the function $\hat{E}^{(i)}_k(x[n],y[n])$ in \eqref{eqn:hat:E} is jointly concave with respect to $x[n]$ and $y[n]$, and therefore, the objective function in problem \eqref{eqn:iteration} is jointly concave with respect to $\{x[n],y[n]\}$. As a result, problem \eqref{eqn:iteration} is a convex optimization problem, and thus can be optimally solved by standard convex optimization techniques such as the interior point method \cite{Boyd:Book}. Furthermore, as shown in Lemma \ref{lemma:4.3}, the objective function in problem \eqref{eqn:iteration} serves as a lower bound for that in problem \eqref{eqn:speed:discretized:P3}. Therefore, after each iteration $i$, the objective function of problem \eqref{eqn:speed:discretized:P3} achieved by $\{x^{(i)}[n],y^{(i)}[n]\}$ monotonically increases \cite{ZengZhangLim2016b}. As problem \eqref{eqn:speed:discretized:P3} has a finite optimal value, the SCP-based algorithm in \eqref{eqn:iteration} will converge to a locally optimal solution to problem \eqref{eqn:speed:discretized:P3} in general.

It is worth noting that the performance of the SCP-based algorithm depends on the choice of the initial trajectory $\{x^{(0)}[n],y^{(0)}[n]\}$. Here, we choose the discretized version of the proposed successive hover-and-fly trajectory obtained in Algorithm 2 as $\{x^{(0)}[n],y^{(0)}[n]\}$. In this case, the SCP-based trajectory design can always achieve a performance no worse than the successive hover-and-fly trajectory design, as will be validated by the numerical results later.

\section{Numerical Results}\label{sec:numerical}

In this section, we provide numerical results to evaluate the performance of our proposed trajectory designs. In the simulation, we set $\beta_0= -30$ dB, $H = 5$ m, and $P = 40$ dBm. For all simulations given below, we consider the average received power by the ER, which is obtained by normalizing the total received energy by the charging duration $T$.

\subsection{Sum-Energy Maximization}

\begin{figure*}
\begin{minipage}[t]{0.49\linewidth}
\centering
\includegraphics[width=\textwidth]{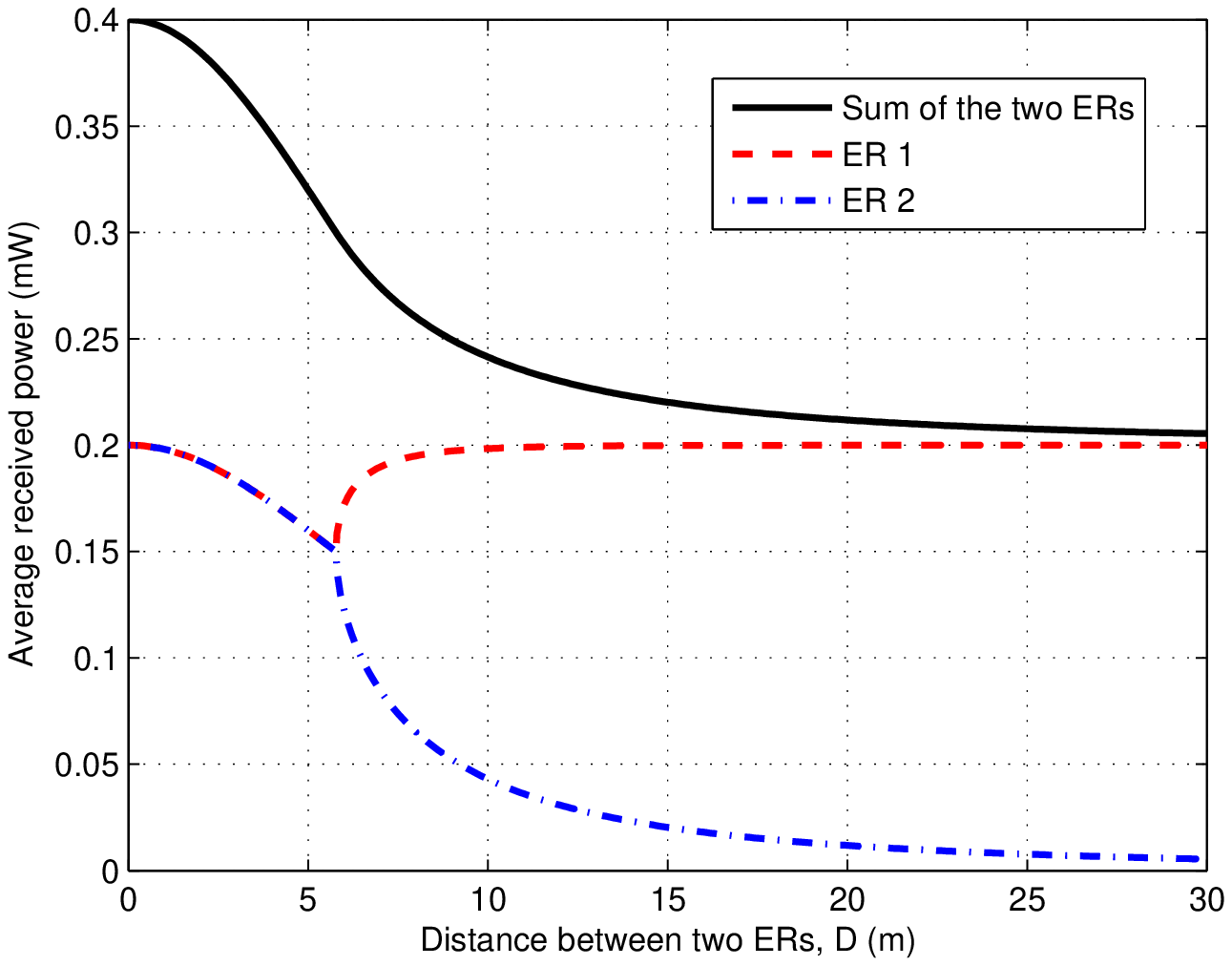}
\caption{The average received power with the sum-energy maximization versus the distance $D$ between the two ERs.} \label{fig2ERSumEnergy}\vspace{-0em}
\end{minipage}
\hfill
\begin{minipage}[t]{0.49\linewidth}
\centering
\includegraphics[width=\textwidth]{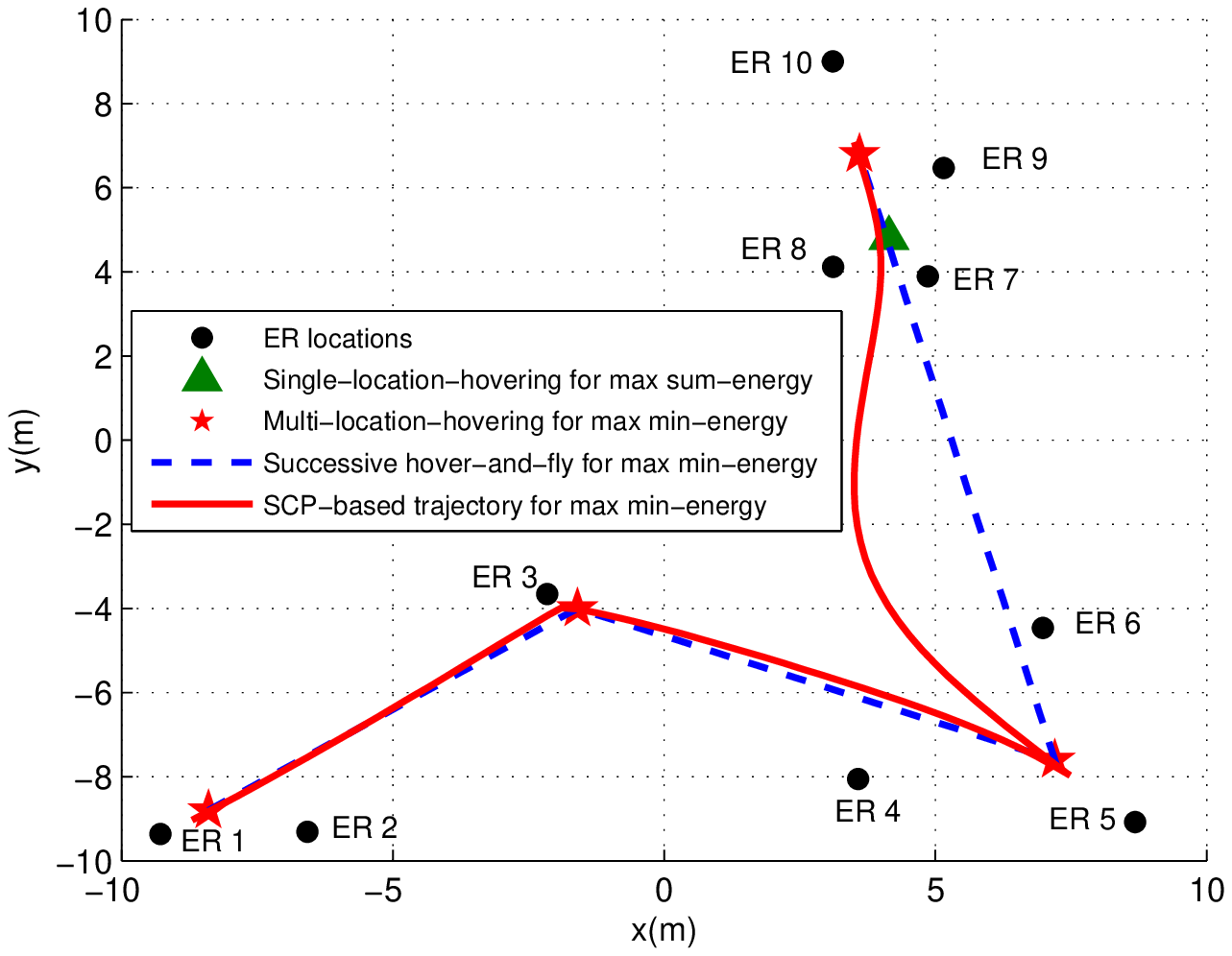}
\caption{Trajectory designs for a UAV-enabled WPT system with $K = 10$ ERs.} \label{fig:ten:user:trajectory}\vspace{-2em}
\end{minipage}
\end{figure*}

%

This subsection evaluates the performance of our proposed optimal solution for the sum-energy maximization problem (P1). First, we consider the case with $K = 2$ ERs, where $x_1 = -D/2, x_2 = D/2$, and $y_1=y_2=0$, with $D$ denoting the distance between the two ERs. Note that Proposition \ref{proposition:3.2} indicates that when $D > 2H/\sqrt{3} = 5.77$ m, there are two optimal single-location-hovering solutions to (P1). In the simulation, we choose the one with $x^\star(t) = - \xi, y^\star(t) = 0, \forall t \in \mathcal T$ (with $\xi$ given in \eqref{eqn:xi}), such that the single hovering location is closer to ER 1 than ER 2. Fig. \ref{fig2ERSumEnergy} shows the average received power versus $D$. When $0\le D\le 2H/\sqrt{3}$, it is observed that the average received power by ER 1 or ER 2 is identical. This is because in this case, the optimal solution is obtained by letting the UAV hover above the middle point between the two ERs during the whole charging period (see Proposition \ref{proposition:3.2}). On the other hand, when $D > 2H/\sqrt{3}$, it is observed that as $D$ increases, the average received power by ER 1 (the ER closer to the UAV) increases, while that by ER 2 (the ER farther away from the UAV) decreases. This is due to the fact that as $D$ increases in this case, the optimal hovering location (selected in this example) becomes closer to ER 1. Accordingly, the average sum received power of the two ERs is dominated by the received power of ER 1, and the near-far fairness issue becomes more severe as $D$ becomes larger.


\begin{figure}
\centering
 \epsfxsize=1\linewidth
    \includegraphics[width=8cm]{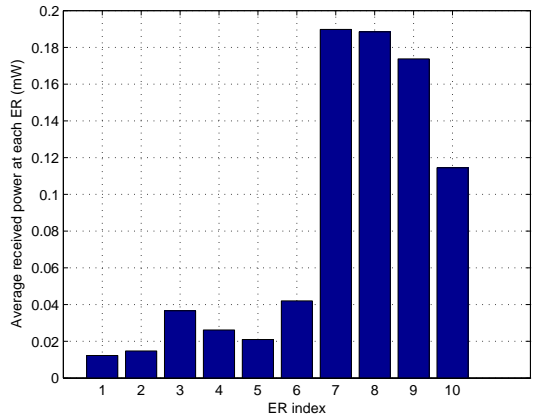}
\caption{The average received power by different ERs with the optimal sum-energy maximization trajectory design.} \label{fig:ten:user:D10}\vspace{-2em}
\end{figure}

Next, we consider a UAV-enabled WPT system with $K = 10$ ERs, whose locations are shown in Fig. \ref{fig:ten:user:trajectory}. In this figure, the green triangle indicates the optimal hovering location for sum-energy maximization (i.e., $(x^\star,y^\star,H)$ given in \eqref{eqn:x_star:y_star}). It is observed that this hovering location is close to ERs 7-10, but far away from other ERs, especially ER 1. Fig. \ref{fig:ten:user:D10} shows the corresponding average received power by each individual ER. It is observed that ERs 7-10 receive much higher energy than the other ERs, which illustrates the near-far fairness issue in the sum-energy maximization for this multiuser WPT system with more than two ERs.

\vspace{-1em}

\subsection{Min-Energy Maximization}

\begin{figure*}
\begin{minipage}[t]{0.49\linewidth}
\centering
\includegraphics[width=\textwidth]{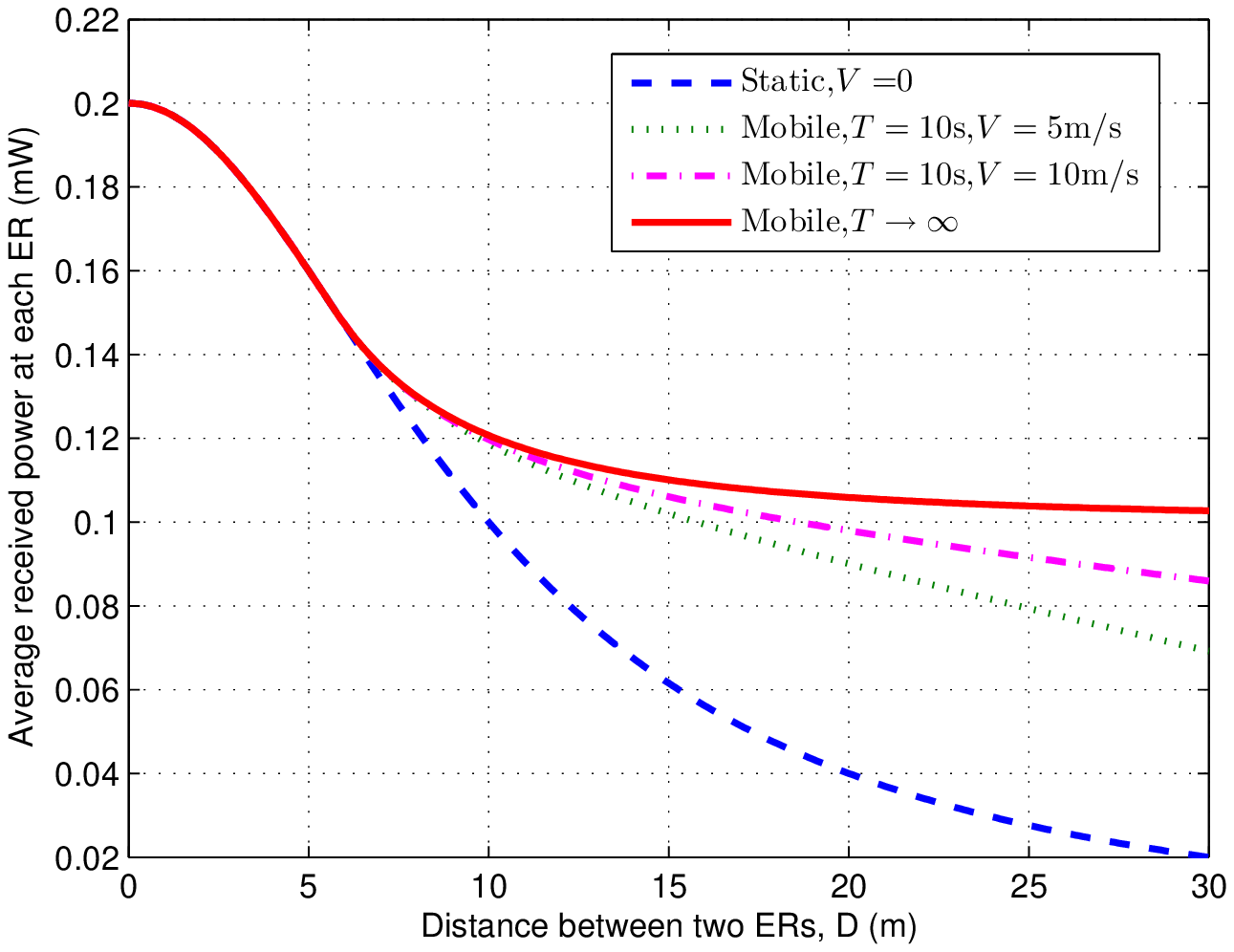}
\caption{The average received power at each ER with the min-energy maximization trajectory design with $K=2$ ERs.} \label{fig2ERMinEnergy}\vspace{-2em}
\end{minipage}
\hfill
\begin{minipage}[t]{0.49\linewidth}
\centering
\includegraphics[width=\textwidth]{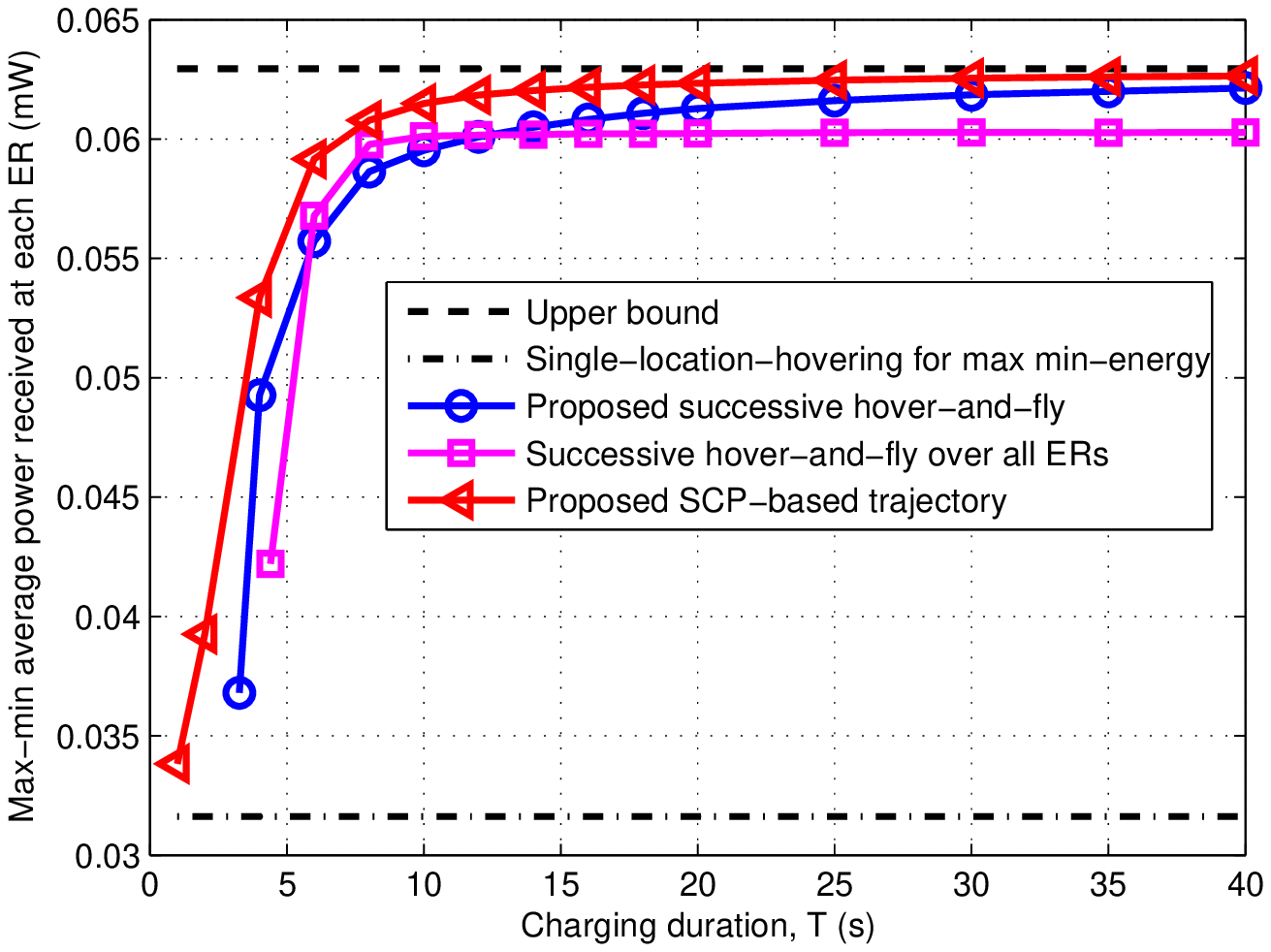}
\caption{The max-min average received power at each ER versus the charging duration $T$ with $K = 10$ ERs.} \label{fig:five:user:D10:common}\vspace{-2em}
\end{minipage}
\end{figure*}

In this subsection, we evaluate the performance of our proposed trajectory designs for the min-energy maximization problem (P2). First, we consider the case with $K=2$ ERs. Fig. \ref{fig2ERMinEnergy} shows the average received power by each ER with the optimal trajectory obtained by solving problem (P2-2ER), with different values of the maximum UAV speed $V$ and the charging duration $T$. It is observed that when $D \le  2H/\sqrt{3} = 5.77{\text{m}}$, the static UAV design (with $V=0$) achieves the same performance as the mobile UAV design, regardless of $V$ and $T$. By contrast, when $D > 5.77$m, the proposed mobile UAV design achieves higher received power than the static UAV design (with the UAV fixed at $(0,0,H)$), and the performance gain becomes more pronounced as $D$ increases. Furthermore, it is observed that as the UAV's maximum speed $V$ increases, the average max-min received power by the ERs increases, as the traveling time between the two ERs becomes less significant.

Next, we consider the min-energy maximization problem for the UAV-enabled WPT system in Fig. \ref{fig:ten:user:trajectory} with $K=10$ ERs. Note that besides the single-location-hovering solution for the sum-energy maximization problem (P1), Fig. \ref{fig:ten:user:trajectory} also shows the multi-location-hovering solution for problem (P3) without the UAV speed constraint, as well as the proposed successive hover-and-fly trajectory design and the SCP-based trajectory design for problem (P2) with the UAV speed constraint considered, by assuming $T = 20$s. First, it is observed that there are $\Gamma = 4$ optimal hovering locations for the min-energy maximization problem (P3), which are close to ERs 1-2, ER 3, ERs 4-6, and ERs 7-10, respectively. This clearly shows that when the ERs are close to each other (e.g., ERs 7-10), then the UAV should hover above one single location above them for charging them more efficiently. It is also observed that the SCP-based trajectory design and the successive hover-and-fly trajectory design both visit the $\Gamma$ optimal hovering locations. Furthermore, it is observed that the SCP-based trajectory in general deviates from the successive hover-and-fly trajectory when flying from one hovering location to another.

Fig. \ref{fig:five:user:D10:common} shows the max-min average power received by all ERs in the 10-ER WPT system in Fig.  \ref{fig:ten:user:trajectory} versus the charging duration $T$,  for our proposed trajectory designs, as compared to the following two benchmark schemes.
\begin{itemize}
  \item {\it Single-location hovering for minimum energy maximization:} During the whole charging period, the UAV hovers at a fixed location $(x_{\text{fix}},y_{\text{fix}},H)$ obtained in \eqref{eqn:fix}.
  \item {\it Successive hover-and-fly over all ERs:} In this successive hover-and-fly scheme, instead of optimizing the UAV's hovering locations, they are simply set as the $K$ locations above the $K$ ERs. This scheme can be implemented by using Algorithm 2 via replacing $\{(x^*_\gamma,y^*_\gamma,H)\}_{\gamma=1}^{\Gamma}$ as  $\{(x_k,y_k,H)\}_{k=1}^K$.
\end{itemize}
In Fig. \ref{fig:five:user:D10:common}, the upper bound corresponds to the optimal value achieved by (P3) with the UAV's maximum speed constraints ignored. It is observed that the two proposed trajectory designs, namely the successive hover-and-fly and the SCP-based trajectory designs, outperform the single-location-hovering design, and achieve higher average max-min average power as $T$ becomes large. When $T\ge 15$ s, the proposed successive hover-and-fly and the SCP-based trajectory designs also outperform the successive hover-and-fly trajectory over all the ERs. Furthermore, it is observed that the SCP-based trajectory achieves better performance than the successive hover-and-fly trajectory, and converges  to the upper bound, when $T$ becomes large.\vspace{-1em}

%

\section{Conclusion}\label{sec:conclusion}

This paper studies a new UAV-enabled multiuser WPT system. We exploit the mobility of the UAV to maximize the energy transferred to all ERs over a given charging period by optimizing the UAV's trajectory under its practical speed constraint. First, we consider the sum-energy maximization of all ERs and obtain the optimal solution to this problem, which shows that the UAV should hover at only one fixed location during the whole charging period. However, this single-location-hovering solution may lead to unfair performance among ERs due to their different distances from the UAV. To achieve fairness among all ERs, we consider an alternative problem to maximize the minimum energy transferred to all ERs. We first consider the relaxed problem by ignoring the UAV speed constraints and derive the optimal solution, which shows that the UAV should hover over multiple fixed locations with optimal hovering time allocations among them. Based on this solution, we further propose two new trajectory designs for solving the min-energy maximization problem in the general case with the UAV speed constraint considered. Numerical results show that the proposed UAV-enabled WPT system with optimized UAV trajectory significantly enhances the WPT performance over the conventional WPT system with fixed ETs, and yet achieves fair energy delivery to ERs. In future work, we will extend our study to the general multi-UAV scenario with joint trajectory design for  multiple UAVs to optimize their cooperative WPT performance.
\vspace{-1em}

\appendix

\subsection{Proof of Proposition \ref{proposition:3.1}}\label{appendix:A}

First, we consider a relaxed problem of (P1) in the ideal case by ignoring the speed constraints in \eqref{eqn:UAV:speed:constraint}. By using (\ref{eqn:sum:energy}), the relaxed problem can be written as
\begin{align}
\max_{\{x(t),y(t)\}} \int_{0}^T \psi(x(t), y(t)) \text{d}t.\label{eqn:problem:P1:Upper}
\end{align}
Due to the relaxation of the constraints \eqref{eqn:UAV:speed:constraint}, the optimal value of problem \eqref{eqn:problem:P1:Upper} serves as an upper bound for that of (P1). It is evident that problem (\ref{eqn:problem:P1:Upper}) can be decomposed into different subproblems as follows:
\begin{align}\label{eqn:problem:sub}
\max_{x(t),y(t)}  \psi(x(t), y(t)), \forall t\in\mathcal T,
\end{align}
where each subproblem corresponds to a time instant $t$. By comparing \eqref{eqn:x_star:y_star} and \eqref{eqn:problem:sub}, it is evident that $x(t) = x^\star$ and $y(t) = y^\star$ is the optimal solution to problem \eqref{eqn:problem:sub} for any $t\in\mathcal T$. Therefore, $\{x^\star(t),y^\star(t)\}$ given in \eqref{eqn:optimal:solution:P1} is indeed the optimal solution to problem \eqref{eqn:problem:P1:Upper}.

Next, it is evident that $\{x^\star(t),y^\star(t)\}$ given in \eqref{eqn:optimal:solution:P1} is always feasible to problem (P1), since the optimal location is fixed and thus no UAV flying is needed. Furthermore, the objective value achieved by $\{x^\star(t),y^\star(t)\}$ for problem (P1) is the same as the optimal value for problem \eqref{eqn:problem:P1:Upper}. Since the optimal value of problem \eqref{eqn:problem:P1:Upper} is an upper bound of that of (P1), we can conclude that $\{x^\star(t),y^\star(t)\}$ is indeed the optimal solution to (P1). Therefore, Proposition \ref{proposition:3.1} is proved.

\subsection{Proof of Lemma \ref{lemma:3.1}}\label{appendix:B}

The first property of $\hat{\psi}(x)$ can be easily shown and the detail is thus omitted here. In the following, we prove the second and third properties for $\hat{\psi}(x)$. Note that the first-order derivative of $\hat{\psi}(x)$ can be obtained as
\begin{align*}
  &\hat\psi'(x) = -4\beta_0 P x \frac{x^4 + 2(D^2/4+H^2)x^2 - 3D^4/16 + H^4-H^2D^2/2}{(x^2 + D^2/4 + H^2 -Dx)^2(x^2 + D^2/4 + H^2 + Dx)^2}.
\end{align*}
Then we consider the two cases with $D \le 2H/\sqrt{3}$ and $D > 2H/\sqrt{3}$, respectively.

When $D \le 2H/\sqrt{3}$, it can be shown that $\hat\psi'(x) = 0$ has only one single real solution given by $x = 0$. Furthermore, we have $\hat\psi'(x) > 0$ for any $x < 0$ and $\hat\psi'(x) < 0$ for any $x > 0$. As a result, $\hat\psi(x)$ is monotonically increasing and decreasing over $x \in (-\infty,0)$ and $x \in (0,\infty)$, respectively. Therefore, the second property of $\hat{\psi}(x)$ is proved.

On the other hand, when $D > 2H/\sqrt{3}$, there exist three real solutions to the equation $\hat\psi'(x) = 0$, which are given by $-\xi$, $0$, and $\xi$, respectively. Furthermore, it can be shown that $\hat\psi'(x) > 0$ for $x \in (-\infty,-\xi)$, $\hat\psi'(x) < 0$ for $x \in (-\xi,0)$, $\hat\psi'(x) > 0$ for $x \in (0,\xi)$, and $\hat\psi'(x) < 0$ for $x \in (\xi,\infty)$. As a result, $\hat\psi(x)$ is monotonically increasing, decreasing, increasing, and decreasing over $x\in  (-\infty,-\xi)$, $(-\xi,0)$, $(0,\xi)$, and $(\xi,\infty)$, respectively. Furthermore, as $\hat{\psi}(- \xi) = \hat{\psi}(\xi)$ based on the first property of the function, $x=-\xi$ and $\xi$ correspond to two equivalent maximizers of $\hat\psi(x)$. Therefore, the third property of $\hat{\psi}(x)$ is proved.

By combining the above properties, this lemma is proved.

\subsection{Proof of Proposition \ref{proposition:4.1}}\label{appendix:C:new}

First, it is evident that at the optimal solution $\{x^*(t)\}$ to (P3-2ER), the received energy at the two ERs must be equal. Furthermore, if the UAV stays at a location $(\hat{x},0,H)$ for a certain duration, then it should stay at the symmetric location $(-\hat{x},0,H)$ for the same duration, since otherwise, we can always increase the min-energy of the two ERs by adjusting the two locations to be symmetric. As a result, without loss of optimality, we only need to consider the following symmetric trajectory $\{x(t)\}$ satisfying
\begin{align}\label{eqn:symmetric}
x(t) = - x(T - t) \le 0, \forall t\in [0,T/2).
\end{align}
Based on \eqref{eqn:symmetric}, it follows from (\ref{eqn:hat:Q1}), (\ref{eqn:hat:Q2}), and (\ref{eqn:hat:E}) that the received energy values at the two ERs are identical, i.e.,
\begin{align}\label{eqn:E1:E2:equal}
\hat E_1(\{x(t)\}) = \hat E_2(\{x(t)\}) = \int_{0}^{T/2} \hat{\psi}(x(t))\text{d}t,
\end{align}
where $\hat{\psi}(x(t))$ is given in \eqref{eqn:hat:psi}. Accordingly, we can re-express problem (P3-2ER) as
\begin{align}
\max_{\{{x}(t)\}}  ~& \int_{0}^{T/2} \hat{\psi}(x(t))\text{d}t.\label{eqn:P3:2ER:Reformulation}
\end{align}
Based on Lemma \ref{lemma:3.1}, if $D \le 2H/\sqrt{3}$, then the optimal solution to \eqref{eqn:P3:2ER:Reformulation} is given by $x^*(t) = 0, \forall t \in [0,T/2)$. If $D > 2H/\sqrt{3}$, then the optimal solution is $x^*(t) = -\xi$, $\forall t \in [0,T/2)$. By also choosing $x^*(t) = \xi, \forall t \in [T/2,T)$ due to symmetry and using \eqref{eqn:symmetric}, this proposition is proved.

\subsection{Proof of Proposition \ref{proposition:4}}\label{appendix:D}

For $K=2$, the optimality of the successive hover-and-fly trajectory can be easily proved in the case of $D \le 2H/\sqrt{3}$. Therefore, in the following, we focus on the proof for the case of $D > 2H/\sqrt{3}$.

First, similarly as Proposition \ref{proposition:4.1} for problem (P3-2ER), the optimal trajectory solution to (P2-2ER) should satisfy the symmetric property in \eqref{eqn:symmetric}. In this case, problem (P2-2ER) is re-expressed as follows based on \eqref{eqn:E1:E2:equal}.
\begin{align}
\max_{\{{x}(t)\}}  ~& \int_{0}^{T/2} \hat{\psi}(x(t))\text{d}t \nonumber\\
{\mathrm{s.t.}}~& |\dot{x}(t)| \le V, \forall t \in [0,T/2], \nonumber\\
& x(T/2) = 0.\label{eqn:P3:2ER:Symmetric}
\end{align}
Without loss of optimality, we only need to consider $x(t) \le 0, \forall t \in [0,T/2]$.

Next, based on Lemma \ref{lemma:3.1}, $\hat\psi(x)$ is monotonically increasing over $x\in  (-\infty,-\xi)$ and decreasing over $(-\xi,0)$. As a result, it can be shown that if $\xi \le VT/2$, the UAV should maximize its hovering time at the location $(-\xi,0,H)$, and fly to the middle point $(0,0,H)$ with its maximum speed $V$. In other words, we have $x^{**}(t) = -\xi, \forall t \in [0,T/2-\xi/V]$, and $x^{**}(t) = Vt -VT/2, \forall t \in (T/2-\xi/V,T/2]$. On the other hand, if $\xi > VT/2$, then the UAV should fly with the maximum speed from the location $(- VT/2,0,H)$ to the middle point $(0,0,H)$ to maximize the objective function in \eqref{eqn:P3:2ER:Symmetric}, i.e., $x^{**}(t) = -VT/2 + Vt, \forall t \in [0,T/2]$.

Combining the above result with the optimal symmetric trajectory in \eqref{eqn:symmetric}, this proposition is thus proved.

\subsection{Proof of Lemma \ref{lemma:4.3}}\label{appendix:E}

Define a function $g_1(z) = \frac{\beta_0P \Delta}{z +H^2}$ with $H^2 > 0$, which is convex with respect to $z \ge 0$. As the first-order Taylor expansion of a convex function is a global under-estimator of the function values \cite{Boyd:Book}, for any given $z_0 \ge 0$, it follows that $g_1(z) \ge g_1(z_0) + g_1'(z_0)(z - z_0)$, or equivalently,
\begin{align}\label{eqn:aa}
\frac{\beta_0P \Delta}{z +H^2} \ge \frac{\beta_0P \Delta}{z_0 +H^2} - \frac{\beta_0P \Delta}{(z_0 +H^2)^2} (z - z_0).
\end{align}
For any given $k\in\mathcal K,n \in \mathcal N, i\ge 0$, by substituting $z = (x[n] - x_k)^2 + (y[n] - y_k)^2$ and $z_0 = (x^{(i)}[n] - x_k)^2 + (y^{(i)}[n] - y_k)^2$ into \eqref{eqn:aa}, then \eqref{eqn:lower:bound} follows. Furthermore, note that the equality holds for \eqref{eqn:aa} for $z = z_0$, and therefore, the equality in \eqref{eqn:strict:eqality} holds. This lemma is thus proved.

\end{document}